\documentclass[A4,12pt,fleqn]{article}

\usepackage{amsmath,cite,epsfig,multicol}

\def\a               {\alpha}
\def\b               {\beta}
\def\d               {\delta}

\def\t               {\theta}

\def\x               {\chi}

\def\L               {{\cal L}}

\def\P               {{\cal P}}

\def\ti              {\tilde}

\def\nt              {\ti\x^0}
\def\ch              {\ti\x^\pm}
\def\sf              {\ti f}
\def\stau            {\ti \tau}
\def\stop            {\ti t}
\def\sbottom         {\ti b}
\def\staun           {\ti \nu_\tau}
\def\tsf             {\theta_{\ti f}}

\def\phmu            {\phi_\mu}

\def\phsf            {\varphi_{\!\sf}^{}}
\def\phst            {\varphi_{\ti t}^{}}
\def\phsb            {\varphi_{\ti b}^{}}
\def\phstau          {\varphi_{\ti\tau}^{}}

\newcommand{\mnt}[1] {m_{\ti \x^0_{#1}}}
\newcommand{\mch}[1] {m_{\ti \x^\pm_{#1}}}
\newcommand{\msf}[1] {m_{\sf_{#1}}}
\newcommand{\mst}[1] {m_{\ti t_{#1}}}
\newcommand{\msb}[1] {m_{\ti b_{#1}}}

\def\PL              {P_L^{}}
\def\PR              {P_R^{}}

\def\to              {\rightarrow}

\newcommand{\nn}{\nonumber}
\newcommand{\noi}{\noindent}

\newcommand{\mbf}      {\boldmath}
\newcommand{\sfrac}[2] {{\textstyle \frac{#1}{#2}}}
\newcommand{\smaf}[2]  {{\textstyle \frac{#1}{#2} }}
\newcommand{\pif}      {\smaf{\pi}{2}}

\newcommand{\eq}[1]  {\mbox{(\ref{eq:#1})}}
\newcommand{\fig}[1] {\mbox{Fig.~\ref{fig:#1}}}
\newcommand{\figs}[1] {\mbox{Figs.~\ref{fig:#1}}}
\newcommand{\Fig}[1] {\mbox{Figure~\ref{fig:#1}}}

\renewcommand{\Re}{{\cal R}e\,}
\renewcommand{\Im}{{\cal I}m\,}

\newcommand{\gsim}{\;\raisebox{-0.9ex}
           {$\textstyle\stackrel{\textstyle >}{\sim}$}\;}
\newcommand{\lsim}{\;\raisebox{-0.9ex}{$\textstyle\stackrel{\textstyle<}
           {\sim}$}\;}

\begin{document}

\vspace*{-18mm}
\begin{flushright}
CERN-PH-TH/2004-086\\
HEPHY-PUB 790/04\\
IISC-CHEP/7/04\\
FI2004-15\\
hep-ph/0405167
\end{flushright}

\vspace*{6mm}

\begin{center}

{\LARGE\bf Fermion polarization in sfermion decays as\\[3mm]
           a probe of CP phases in the MSSM}\\[8mm]

{\large Thomas~Gajdosik$^{\,1}$, Rohini~M.~Godbole$^{\,2}$,
        Sabine~Kraml$^{\,3,4}$}\\[6mm]

{\it
$^1$~Institute of Physics, Vilnius LT-2600, Lithuania\\[2mm]
$^2$~Centre for High Energy Physics, Indian Institute of Science,
     Bangalore 560012, India\\[2mm]
$^3$~Inst.\ f.\ Hochenergiephysik, \"Osterr.\ Akademie d.\ Wissenschaften,
     1050 Vienna, Austria\\[2mm]
$^4$~Department of Physics, CERN, Theory Division,
     1211 Geneva 23, Switzerland\\[2mm]
}

\end{center}

\begin{abstract}
\noindent
  The longitudinal polarization of fermions (tops and taus)
  produced in sfermion decays to neutralinos or charginos can be a
  useful tool for the determination of SUSY parameters. We discuss
  this fermion polarization in the context of the MSSM with
  complex parameters. We show that the dependence on CP-violating
  phases can be large and that the fermion polarization may hence
  be used as a sensitive probe of CP phases in the MSSM.
\end{abstract}

\section{Introduction}

CP violation, initially observed~\cite{kaon} only in the
$K_0$--$\bar {K_0}$ system, is one feature of the Standard Model
(SM) that still defies clear theoretical understanding. The CKM
picture, which describes {\it all} the {\it observed} CP violation
in terms of a single phase in the quark-mixing matrix, has been
vindicated by the recent measurements of $B_0$--$\bar {B_0}$
mixing at BELLE and BABAR~\cite{bellebabar}.
CP violation is in fact one of the necessary ingredients for
generating the observed excess of baryons over antibaryons in the
Universe \cite{Sakharov:dj,Dolgov:2002kw}. The amount of CP
violation present in the quark sector is, however, too small to
generate a baryon asymmetry of the observed level of $N_B
/N_{\gamma} \simeq 6.1 \times 10^{-10}$~\cite{Bennett:2003bz}. New
sources of CP violation {\it beyond} the SM are
therefore a necessity~\cite{Dine:2003ax}.

Supersymmetry (SUSY) is arguably the most attractive extension of
the SM, as it solves, for instance, the problem of the instability
of the electroweak symmetry-breaking scale against radiative
corrections. Already the Minimal Supersymmetric Standard Model
(MSSM) \cite{Haber:1984rc} provides possible new sources of CP
violation through additional CP-violating phases, which cannot be
rotated away by simple field redefinitions. A large number of
these phases, particularly those involving sparticles of the first
and to a large extent of the second generation, are severely
constrained by measurements of the electric dipole moments (EDMs)
of the electron, muon, neutron as well as ${}^{199}$Hg and
$^{205}$Tl. However, these constraints are model-dependent. It has
been demonstrated
\cite{Ibrahim:1997nc,Brhlik:1998zn,Bartl:1999bc,Falk:1998pu,Falk:1999tm}
that cancellations among different diagrams allow certain
combinations of these phases to be large in a general MSSM.
Furthermore, if the sfermions of the first two generations are
sufficiently heavy, above the 1 TeV range, the EDM constraints on
the phase of the higgsino mass parameter $\mu=|\mu|e^{i\phi_\mu}$,
in general constrained to $\phi_\mu^{}\lsim 10^{-2}$, get weaker;
the sfermions of the third generation can still be light.
   Non-vanishing phases of $\mu$ and/or the trilinear scalar couplings
$A_{t,b}$ can induce explicit CP violation in the Higgs sector via
loop corrections
\cite{Pilaftsis:1998pe,Demir:1999hj,Choi:2000wz,Carena:2000yi}.
Though these phases generate EDMs independently of the first two
generations of sfermions, the EDMs are suppressed by the mass
scale of the two heavy Higgses
\cite{Chang:1998uc,Pilaftsis:1999td}. For a thorough discussion of
the EDMs see \cite{Demir:2003js} and references therein.
The above mentioned phases can also have a significant influence
on the Higgs production rates in the gluon fusion mode at the
Tevatron and the LHC \cite{Dedes:1999sj,Choi:2001iu}. MSSM CP
phases can hence change the Higgs phenomenology at colliders quite
substantially.

All this makes the MSSM with CP-violating phases a very attractive
proposition. It has therefore been the subject of many recent
investigations, studying the implications of these phases on
neutralino/chargino production and decay
\cite{Choi:1998ei,Choi:1999mv,Kneur:1999nx,Choi:2000ta,
      Choi:2000kt,Choi:2001ww,Barger:2001nu,Bartl:2003tr,
      Bartl:2003gr,Choi:2003pq,Bartl:2004ut,Choi:2004rf},
on the third generation of sfermions
\cite{Bartl:2002hi,Bartl:2002uy,Bartl:2002bh,
      Bartl:2003ck,Bartl:2003he,Bartl:2003pd},
as well as the neutral
\cite{Choi:2001pg,Arhrib:2001pg,Choi:2002zp,Carena:2002bb,Borzumati:2004rd}
and charged \cite{Christova:2002ke} Higgs sector. Various CP-even
and CP-odd (T-odd) observables, which can give information on these
phases, have been identified. It is interesting to note that
CP-even observables such as masses, branching ratios, cross
sections, etc., often afford more precise probes thanks to the
larger magnitude of the effects. For direct evidence of CP violation,
however, CP-odd/T-odd observables as discussed e.g.\ in
\cite{Choi:1999mv,Choi:2002zp,Bartl:2003ck,Bartl:2003tr,Bartl:2003gr,
      Choi:2003pq,Bartl:2004ut,Choi:2004rf}
have to be measured.

The latest study of the $\stop, \sbottom$ sector in \cite{Bartl:2003pd}
demonstrates that it may be possible to determine the real and imaginary
parts of $A_t$ to a precision of 2--3\% from a fit of the MSSM Lagrange
parameters to masses, cross sections and branching ratios at a future
$e^+e^-$ Linear Collider (LC).
This requires that both the $\ti t_{1,2}^{}$, $\ti b_{1,2}^{}$ mass
eigenstates can be produced at the LC and the branching ratios measured
with high precision. In the $\stau/\staun$ sector
\cite{Bartl:2002bh,Bartl:2002uy} the precision on $A_\tau$ is worse,
around 10--20\% for low $\tan\b$ and about 3--7\% for large $\tan\b$.

In this paper, we show that the longitudinal polarization of
fermions produced in sfermion decays, i.e.\ $\ti f\to f\nt$ and
$\ti f\to f'\ch$ with $f(\ti f)$ a third generation (s)quark or
(s)lepton, can also be used as a probe of CP phases. The fermion
polarization can give complementary information to the decay
branching ratios and will in particular be useful if the branching
ratios cannot be measured with high enough precision or if one
decay channel dominates.

The average polarization of fermions produced in sfermion decays
carries information on the $\ti f_L^{}$--$\ti f_R^{}$ mixing as
well as on the gaugino--higgsino mixing \cite{Nojiri:1994it}. The
polarizations that can be measured are those of top and tau; both
can be inferred from the decay lepton distributions. It is its
large mass that causes the $t$ to decay before hadronization and
thus the decay products can carry information about its
polarization. For taus, also the energy distribution of the decay
pions can be used. The polarization of the decay fermions has been
used for studies of MSSM parameter determination in the
CP-conserving case in \cite{Nojiri:1994it,Nojiri:1996fp,Boos:2003vf}.
For the CP-violating case, the phase dependence of the longitudinal
fermion polarization has been mentioned in \cite{Bartl:2002bh}.
We extend these studies by discussing in detail the sensitivity
of the fermion polarization to CP-violating phases in the MSSM.

The paper is organized as follows: in Section 2, we summarize our
notation for the description of the sfermion, neutralino and
chargino systems in the MSSM with CP violation. In Section~3, we
discuss fermion polarization in sfermion decays to neutralinos,
$\sf\to f\,\nt$ with $f=t,\tau$. We present numerical results on
the polarization as a function of different MSSM parameters and
discuss the sensitivity to CP-violating phases in the sfermion and
neutralino sectors. In Section~4 we perform an analogous analysis
for $\sf \to f'\,\ch$ decays. In Section~5 we summarize the
results and present our conclusions.

\section{Notation and conventions}

\subsection{Sfermion system}

Ignoring intergenerational mixing, the sfermion mass matrices can
be written as a series of $2\!\times\!2$ matrices, each of which
describes sfermions of a specific flavour:
\begin{equation}
  {\cal M}_{\sf}^2 =
  \left( \begin{array}{cc}  \msf{L}^2 & a_f^* m_f \\
                            a_f m_f   & \msf{R}^2
  \end{array}\right) \;=\;
  (R^{\sf})^\dagger \left( \begin{array}{cc} \msf{1}^2 & 0 \\
                                            0         & \msf{2}^2
                   \end{array}\right) R^{\sf}
\label{eq:msfmat}
\end{equation}
with
\begin{eqnarray}
  \msf{L}^2 &=& M^2_{\ti L}
    + m_Z^2 \cos 2\beta\,(I_{3L}^f - e_f\sin^2\t_W) + m_f^2, \\[2mm]
  \msf{R}^2 &=& M^2_{\ti R}
    + e_f\,m_Z^2 \cos 2\b\,\sin^2\t_W + m_f^2, \\[1mm]
  a_f &=& A_f - \mu^*\, \{ \cot\b , \tan\b \}
          = |a_f|\, e^{i\phsf} \,,
  \label{eq:aq}
\end{eqnarray}
for $\{$up, down$\}$-type sfermions; $m_f$, $e_f$ and $I_{3}^f$
are the mass, electric charge and the third component of the weak
isospin of the partner fermion, respectively; $M_{\ti L}$,
$M_{\ti R}$ and $A_f$ are soft SUSY-breaking parameters for each
family, and $\mu$ is the higgsino mass parameter; $A_f$ and $\mu$
can have complex phases:
$A_f = |A_f| \, e^{i\phi_{A_{f}}}$ and $\mu = |\mu| \, e^{i\phmu}$.\\

\noi
According to eq.~\eq{msfmat}, ${\cal M}_{\sf}^2$ is diagonalized by a
unitary rotation matrix $R^{\sf}$.
The weak eigenstates $\sf_L^{}$ and $\sf_R^{}$ are thus
related to their mass eigenstates $\sf_1^{}$ and $\sf_2^{}$ by
\begin{equation}
  {\sf_1^{} \choose \sf_2^{}} = R^{\sf}\,{\sf_L^{} \choose \sf_R^{}},
  \hspace{8mm}
  R^{\sf} = \left(\begin{array}{cc}
                     \cos\tsf\;e^{i\phsf} & \sin\tsf \\
                    -\sin\tsf             & \cos\tsf\;e^{-i\phsf}
            \end{array}\right) ,
\label{eq:Rsf}
\end{equation}
with $\tsf$ and $\phsf={\rm Arg}(a_f)$ the sfermion mixing angle
and phase. Since the off-diagonal element of ${\cal M}_{\sf}^2$ is
proportional to $m_f$, this mixing is mostly relevant to the third
generation, $\ti f=\ti t,\ti b,\ti\tau$, on which we concentrate
in the following. The mass eigenvalues are given by
\begin{equation}
  m^2_{\sf_{1\!,2}} = \frac{1}{2} \left( \msf{L}^2 + \msf{R}^2
  \mp \sqrt{(\msf{L}^2 - \msf{R}^2)^2 + 4\, |a_f m_f|^2 } \,\right).
\label{eq:sfmasseigenvalues}
\end{equation}
By convention, we choose $\sf_1^{}$ to be the lighter mass
eigenstate, $\msf{1}\leq\msf{2}$. Notice also that
$\msf{1}\leq\msf{L,R}\leq\msf{2}$. For the mixing angle $\tsf$ we
choose
\begin{equation}
  \cos\tsf = \frac{-|a_f m_f|}
                  {\sqrt{(\msf{L}^2-\msf{1}^2)^2 + |a_f m_f|^2}} \,,
  \qquad
  \sin\tsf = \frac{\msf{L}^2-\msf{1}^2}
                  {\sqrt{(\msf{L}^2-\msf{1}^2)^2 + |a_f m_f|^2}} \,,
\end{equation}
which places $\tsf$ in the 2nd quadrant of the unit circle. The
$\sf_L$--$\sf_R$ mixing is large if $|\msf{L}^2 - \msf{R}^2| \lsim
|a_f m_f|$, with $|\cos\tsf| > \frac{1}{\sqrt 2}\,$ if
$\msf{L}<\msf{R}$ and $|\cos\tsf| < \frac{1}{\sqrt 2}\,$ if
$\msf{R}<\msf{L}$. Moreover, we see that the phase dependence of
$\msf{1,2}^2$ and $R^{\sf}$ is determined by $\cos(\phi_{A_{f}} +
\phmu)$. This dependence is strongest if $|A_f| \simeq |\mu|
\{\cot\b,\tan\b\}$. This issue will be discussed in more detail in
the numerical analyses of Sections~3 and 4.

\subsection{Neutralino system}

In the basis
\begin{equation}
  \Psi_j^0=\left(-i\lambda ',-i\lambda^3,\psi_{H_1}^0,\psi_{H_2}^0\right)
\end{equation}
the neutralino mass matrix is:
\begin{equation}
  {\cal M}_N =
  \left( \begin{array}{cccc}
  M_1 & 0 & -m_Z\sin\theta_W\cos\beta  & m_Z\sin\theta_W\sin\beta \\
  0 & M_2 &  m_Z\cos\theta_W\cos\beta  & -m_Z\cos\theta_W\sin\beta  \\
  -m_Z\sin\theta_W\cos\beta & m_Z\cos\theta_W\cos\beta   & 0 & -\mu \\
   m_Z\sin\theta_W\sin\beta & - m_Z\cos\theta_W\sin\beta & -\mu & 0
  \end{array}\right) .
\label{eq:ntmassmat}
\end{equation}

\noi
The gaugino mass parameters $M_{1,2}$ and the higgsino mass
parameter $\mu$ can in principle all be complex. The phase of
$M_2$ can be rotated away, which leaves us with two phases in this
sector: $\phi_1$, the phase of $M_1$, and $\phmu$, the phase of $\mu$.\\

\noi The matrix of eq.~\eq{ntmassmat} is diagonalized by the
unitary mixing matrix $N$:
\begin{equation}
  N^*{{\cal M}_N} N^\dagger =
  {\rm diag}(\mnt{1},\,\mnt{2},\,\mnt{3},\,\mnt{4})\,,
\end{equation}
where $\mnt{n}$, $n=1,...,4$, are the (non-negative) masses of the
physical neutralino states. We choose the ordering
$\mnt{1}<....<\mnt{4}$. A concise discussion of the neutralino
sector with complex phases can be found in \cite{Choi:2004rf}.

\subsection{Chargino system}

\noi
The chargino mass matrix is:
\begin{equation}
  {\cal M}_C =
  \left( \begin{array}{cc}
    M_2 &\sqrt 2\, m_W\sin\beta \\
    \sqrt 2\,m_W\cos\beta & \mu
  \end{array}\right) \,.
\end{equation}
It is diagonalized by the two unitary matrices $U$ and $V$,
\begin{equation}
  U^*{\cal M}_C V^\dagger = {\rm diag}(\mch{1},\,\mch{2})\,,
\end{equation}
where $\mch{1,2}$ are the masses of the physical chargino states
with $\mch{1}<\mch{2}$.

\section{\mbf Fermion polarization in $\sf\to f\nt$ decays} \label{sect:Pf}

The sfermion interaction with neutralinos is ($i=1,2$; $n=1,...,4$)
\begin{eqnarray}
  \L_{f\sf\nt}
  &=& g\,\bar f\,( f_{Ln}^{\sf}\PR + h_{Ln}^{\sf}\PL )\,\nt_n\,\sf_L^{} +
      g\,\bar f\,( h_{Rn}^{\sf}\PR + f_{Rn}^{\sf}\PL )\,\nt_n\,\sf_R^{}
      + {\rm h.c.}\nn\\
  &=& g\,\bar f\,( a^{\,\sf}_{in}\PR + b^{\,\sf}_{in}\PL )\,\nt_n\,\sf_i^{}
      + {\rm h.c.}
\end{eqnarray}
where
\begin{eqnarray}
   a^{\,\sf}_{in} &=& f_{Ln}^{\sf}\,R_{i1}^{\sf\,*} +
                      h_{Rn}^{\sf}\,R_{i2}^{\sf\,*},   \label{eq:aik}\\
   b^{\,\sf}_{in} &=& h_{Ln}^{\sf}\,R_{i1}^{\sf\,*} +
                      f_{Rn}^{\sf}\,R_{i2}^{\sf\,*}.   \label{eq:bik}
\end{eqnarray}
The $f_{L,R}^{\sf}$ and $h_{L,R}^{\sf}$ couplings are
\begin{align}
  f_{Ln}^{\,\ti t} &= -\sfrac{1}{\sqrt 2}\,(N_{n2}
                        +\sfrac{1}{3}\tan\theta_W N_{n1}) \,, &
  f_{Ln}^{\,\ti b} &= \sfrac{1}{\sqrt 2}\,(N_{n2}
                        -\sfrac{1}{3}\tan\theta_W N_{n1}) \,, \\
  f_{Rn}^{\,\ti t} &= \sfrac{2\sqrt 2}{3}\,\tan\theta_W N_{n1}^*\,, &
  f_{Rn}^{\,\ti b} &= -\sfrac{\sqrt 2}{3}\,\tan\theta_W N_{n1}^*\,, \\
  h_{Rn}^{\ti t} &= -h_t\, N_{n4} = h_{Ln}^{\ti t*}\,, &
  h_{Rn}^{\ti b} &= -h_b\, N_{k3} = h_{Ln}^{\ti b*}\,
\end{align}
for stops and sbottoms, and
\begin{align}
  f_{Ln}^{\ti\tau} &= \sfrac{1}{\sqrt 2}\,(\tan\theta_W N_{n1}+N_{n2})\,,\\
  f_{Rn}^{\ti\tau} &= -\sqrt{2}\,\tan\theta_W N_{n1}^*\,,\\
  h_{Rn}^{\ti\tau} &= -h_\tau\, N_{n3} = h_{Ln}^{\ti\tau*}\,
\end{align}
for staus, with the Yukawa couplings $h_f$ given by
\begin{equation}
  h_t=\frac{m_t}{\sqrt 2\,m_W\sin\b}\,,\qquad
  h_{b,\tau}=\frac{m_{b,\tau}}{\sqrt 2\,m_W\cos\b}\,.
  \label{eq:yuk}
\end{equation}

\noi
The gaugino interaction conserves the helicity of the
sfermion while the higgsino interaction flips it. In the limit
$m_f\ll\msf{i}$, the average polarization of the fermion coming
from the $\sf_i\to f\nt_n$ decay can therefore be calculated as
\cite{Nojiri:1994it}
\begin{equation}
   \P_f^{} = \frac{Br(\sf_i\to\nt_n f_R)-Br(\sf_i\to\nt_n f_L)}
                  {Br(\sf_i\to\nt_n f_R)+Br(\sf_i\to\nt_n f_L)}
           = \frac{|b_{in}^{\sf}|^2-|a_{in}^{\,\sf}|^2}
                  {|b_{in}^{\sf}|^2+|a_{in}^{\,\sf}|^2} \,.
   \label{eq:Pf}
\end{equation}

\noi
Using eqs.~\eq{aik}, \eq{bik} and \eq{Rsf} as well as
$h_{Ln}^{}=h_{Rn}^{*}$, we obtain, for the $\sf_1^{}\to f\nt_n$
decay (omitting the overall factor $g^2$ and dropping the sfermion
and neutralino indices for simplicity):
\begin{eqnarray}
   |b_{1n}^{}|^2-|a_{1n}^{}|^2
   &=& |h_L^{}\cos\t\,e^{-i\varphi} + f_R^{}\sin\t|^2 -
       |f_L^{}\cos\t\,e^{-i\varphi} + h_L^*\sin\t|^2    \nn\\
   &=& (|h_L|^2-|f_L|^2)\cos^2\t - (|h_L|^2-|f_R|^2)\sin^2\t \nn\\
   & & + \,\sin 2\t\,\big[\,
       \Re (f_R^{}-f_L^{})\,(\Re h_L^{}\cos\varphi+\Im h_L^{}\sin\varphi) \nn\\
   & & \hspace*{14mm}
       +\,\Im (f_R^{}+f_L^{})\,(\Im h_L^{}\cos\varphi-\Re h_L^{}\sin\varphi)
       \,\big] \,.
\end{eqnarray}
We see that the phase dependence of $\P_f^{}$ is the largest for
maximal sfermion mixing ($\tsf=3\pi/4$) and if the neutralino has
both sizeable gaugino and higgsino components. It is, moreover,
enhanced if the Yukawa coupling $h_f$ is large.
Furthermore, $\P_f^{}$ is sensitive to CP violation even if just
one phase, in either the neutralino or the sfermion sector, is
non-zero. In particular, if only $A_f$ and thus only the sfermion
mixing matrix has a non-zero phase, the phase-dependent term
becomes
\begin{equation}
   |b_{1n}^{}|^2-|a_{1n}^{}|^2
   \: \stackrel{\phi_1=\phi_\mu=0}{\sim} \:
   h_L^{}(f_L^{}-f_R^{})\sin2\t\cos\varphi \,.
\end{equation}
If, on the other hand, only $\phi_1$ is non-zero we get
\begin{equation}
   |b_{1n}^{}|^2-|a_{1n}^{}|^2
   \; \stackrel{\phi_A=\phi_\mu=0}{\sim} \;
   \big[ \Re h_L^{}\,\Re (f_R^{}-f_L^{}) +
         \Im h_L^{}\,\Im (f_R^{}+f_L^{}) \big]\,\sin 2\t \,.
\end{equation}

    The polarization $\P_{\!f}^{}$, eq.~\eq{Pf}, depends only on
couplings but not on masses. For the numerical analysis we
therefore use $M_1$, $M_2$, $\mu$, $\tan\beta$, $\tsf$ and $\phsf$
as input parameters, assuming $\phmu \approx 0$ to satisfy EDM
constraints more easily: assuming cancellations for the 1-loop
contributions and the CP-odd Higgs mass parameter $m_{A}
> 300$~GeV, 1-loop and 2-loop contributions to the electron EDM (eEDM),
as well as their sum, stay below the experimental limit. We use
the formulae of \cite{Chang:1998uc} for the 2-loop contributions.
    In order not to vary too many parameters, we use,
moreover, the GUT relation $|M_1|=\frac{5}{3}\tan^2\theta_W M_2$
and choose $\tan\b=10$ and $\t_{\ti t}=\t_{\ti\tau}=130^\circ$
(large but not maximal mixing) throughout this section. The free
parameters in our analysis are thus $M_2$, $|\mu|$, and the phases
$\phi_1$, $\phsf$.

Before we present the numerical results, a comment is in order: CP
violation in the neutralino sector is determined by the phases of
$M_1$ and $\mu$, while $\phsf$ originates from relative phases of
$A_f$ and $\mu$. For stops the mixing is dominated by $A_t$, while
for sbottoms and staus it is dominated by $\mu\tan\b$; quite
generally we have $\varphi_{\ti t}\sim \phi_{A_t}$ unless
$|\mu|\sim |A_t|\tan\beta$, and $\varphi_{\ti b,\ti\tau}\sim
-\phmu$ unless $|A_{b,\tau}|\sim |\mu|\tan\beta$. More precisely,
\begin{equation}
  \tan\phsf = \frac{x_f\sin\phi_{A_{f}}+\sin\phmu}{x_f\cos\phi_{A_{f}}
  -\cos\phmu}
  \quad {\rm with} \quad
  x_t = \frac{|A_t|\,\tan\b}{|\mu|}\,, \quad
  x_{b,\tau} = \frac{|A_{b,\tau}|}{|\mu|\tan\b}\,.
\label{eq:xrel}
\end{equation}
For $x_f>1$, any $\phsf$ can be reached by an appropriate choice
of $\phi_{A_f}$, independent of $\phi_\mu$. For $x_f\le 1$,
however, $\phsf$ is restricted by $\phi_\mu$. In the special case
of $x_f=1$ and $\phi_\mu=0$, $\phsf=(\phi_{A_f}+\pi)/2$. In the
stop sector this is not a problem since $x_t>1$ can in general be
easily achieved. For sbottoms and staus, choosing $\varphi_{\ti
b,\ti\tau}$ freely leads, however, to quite large
$|A_{b,\tau}|\sim {\cal O}(|\mu|\tan\b)$, which may in some cases
create problems with charge- or colour-breaking minima.


\begin{figure}[p]
\begin{center} {\setlength{\unitlength}{1mm}
\begin{picture}(148,66)
\put(0,0){\mbox{\epsfig{figure=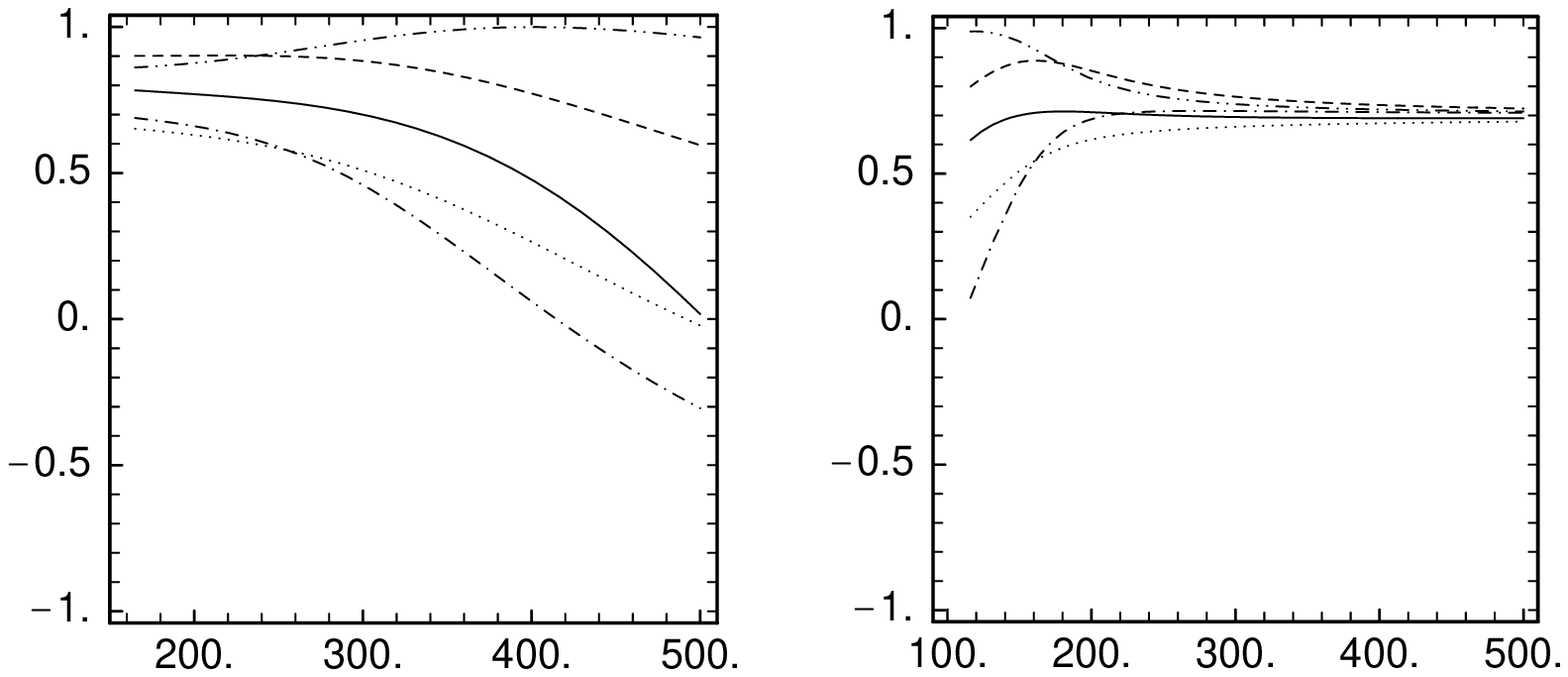,height=7cm}}}
\put(30,0){\mbox{ $M_2$~[GeV] }}
\put(104,0){\mbox{ $|\mu|$~[GeV] }}
\put(-2,37){\mbox{ $\P_{\!\tau}^{}$ }}
\put(72,37){\mbox{ $\P_{\!\tau}^{}$ }}
\put(-2,64){\mbox{ \bf a) }}
\put(72,64){\mbox{ \bf b) }}
\put(16,14){\mbox{ $|\mu|=150$~GeV }}
\put(90,14){\mbox{ $M_2=300$~GeV }}
\end{picture} }
\end{center}
\caption{Average polarization of the tau lepton coming from
$\ti\tau_1\to\tau\nt_1$ decays for $\theta_{\ti\tau}=130^\circ$
and $\tan\beta=10$: in a) as a function of $M_2$ for $|\mu|=150$~GeV,
in b) as a function of $|\mu|$ for $M_2=300$~GeV.
The full, dashed, dotted, dash-dotted, and dash-dot-dotted
lines are for $(\phi_1,\,\varphi_{\ti \tau})=(0,\,0)$,
$(0,\,\frac{\pi}{2})$, $(\frac{\pi}{2},0)$,
$(\frac{\pi}{2},\,\frac{\pi}{2})$, and
$(\frac{\pi}{2},\,-\frac{\pi}{2})$, respectively.
$M_2$ and $\mu$ are taken to be real;
for $|M_1|$ the GUT relation $|M_1|=\frac{5}{3}\tan^2\theta_W M_2$
has been used.
\label{fig:Ptau_M2mu}}
\end{figure}


\begin{figure}[p]
\begin{center} {\setlength{\unitlength}{1mm}
\begin{picture}(148,72)
\put(0,0){\mbox{\epsfig{figure=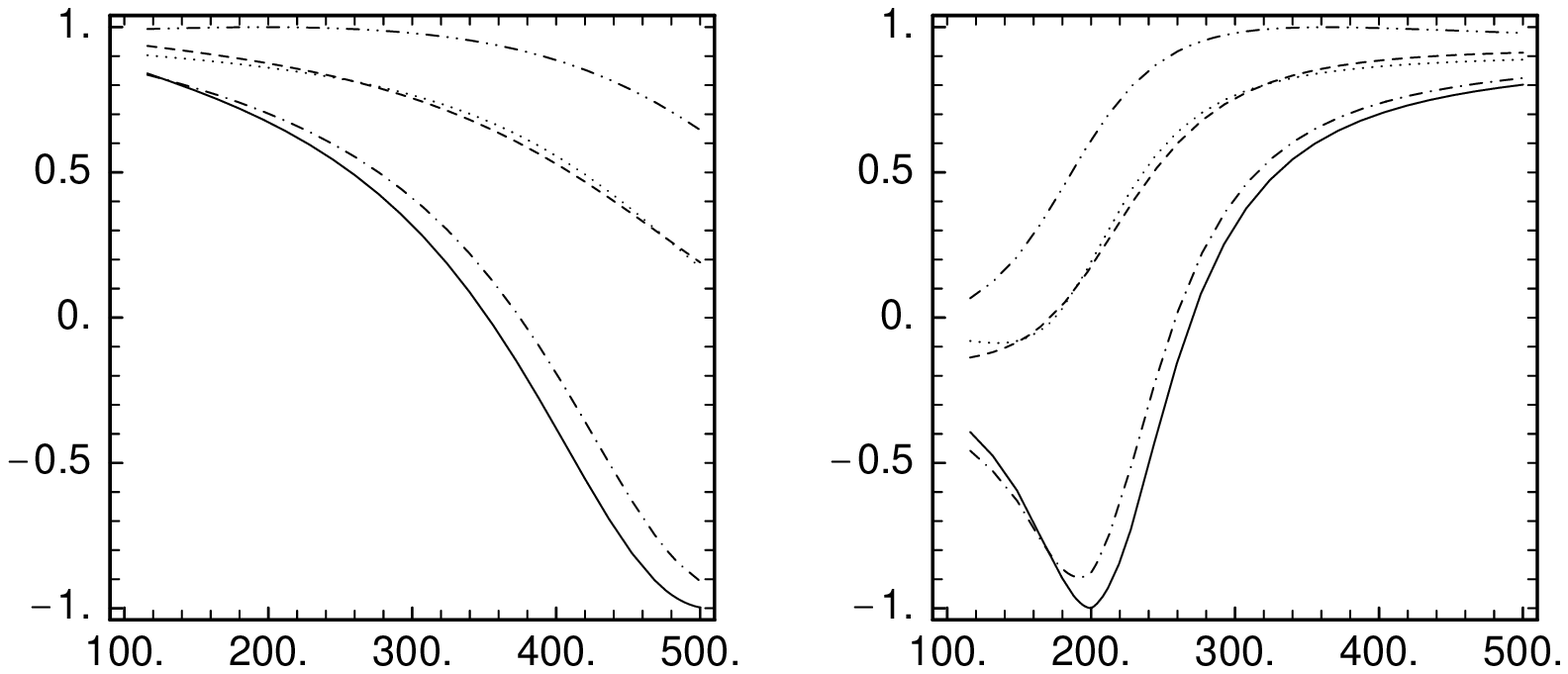,height=7cm}}}
\put(30,0){\mbox{ $M_2$~[GeV] }}
\put(104,0){\mbox{ $|\mu|$~[GeV] }}
\put(-2,37){\mbox{ $\P_{\!t}^{}$ }}
\put(72,37){\mbox{ $\P_{\!t}^{}$ }}
\put(-2,64){\mbox{ \bf a) }}
\put(72,64){\mbox{ \bf b) }}
\put(16,14){\mbox{ $|\mu|=300$~GeV }}
\put(110,14){\mbox{ $M_2=300$~GeV }}
\end{picture} }
\end{center}
\caption{Average polarization of the top quark coming from
$\ti t_1\to t\nt_1$ decays analogous to \fig{Ptau_M2mu}.
It is assumed that $\mst{1}\gg m_t$.
\label{fig:Ptop_M2mu}}
\end{figure}


\Fig{Ptau_M2mu} shows the average tau polarization in
$\ti\tau_1\to\tau\nt_1$ decays as functions of $M_2$ and $|\mu|$
for $\tan\b=10$, $\theta_{\ti\tau}=130^\circ$ and various choices
of $\phi_1$ and $\varphi_{\ti\tau}$. The lower limits of $M_2$ and
$|\mu|$ are given by the LEP2 constraint of $\mch{1}>103$~GeV
\cite{lepsusy}, which automatically takes care of all other LEP
constraints on the gaugino--higgsino sector. As can be seen,
$\P_{\!\tau}^{}$ is quite sensitive to CP phases for $|\mu|<M_2$,
that is if the $\nt_1$ has a sizeable higgsino component.
    Analogously, \fig{Ptop_M2mu} shows the average top polarization
in $\ti t_1\to t\nt_1$ decays. We observe again a strong
dependence on the CP phases if the neutralino has a sizeable
higgsino component. Unlike the case of $\P_{\!\tau}^{}$, for
$\P_{\!t}^{}$ the dependence is still significant when $|\mu|\sim
M_2$. We also note that some phase combinations lead to very
similar polarizations, e.g.\ $\P_{\!t}^{}(\phi_1=0,\phst=0)\sim
       \P_{\!t}^{}(\phi_1=\frac{\pi}{2},\phst=\frac{\pi}{2})$ and
      $\P_{\!t}^{}(\phi_1=0,\phst=\frac{\pi}{2})\simeq
       \P_{\!t}^{}(\phi_1=\frac{\pi}{2},\phst=0)$.
At a future $e^+e^-$ linear collider (LC), 
one expects to be able to measure the tau polarization to about
3--5\% and the top polarization to about 10\% \cite{Boos:2003vf}.
We see from Figs.~\ref{fig:Ptau_M2mu} and \ref{fig:Ptop_M2mu} that
the effects of CP-violating phases may well be visible in
$\P_{\!t}^{}$ and/or $\P_{\!\tau}^{}$, provided $\mu$ is not too
large.

\begin{figure}[p]
\begin{center} {\setlength{\unitlength}{1mm}
\begin{picture}(148,72)
\put(0,0){\mbox{\epsfig{figure=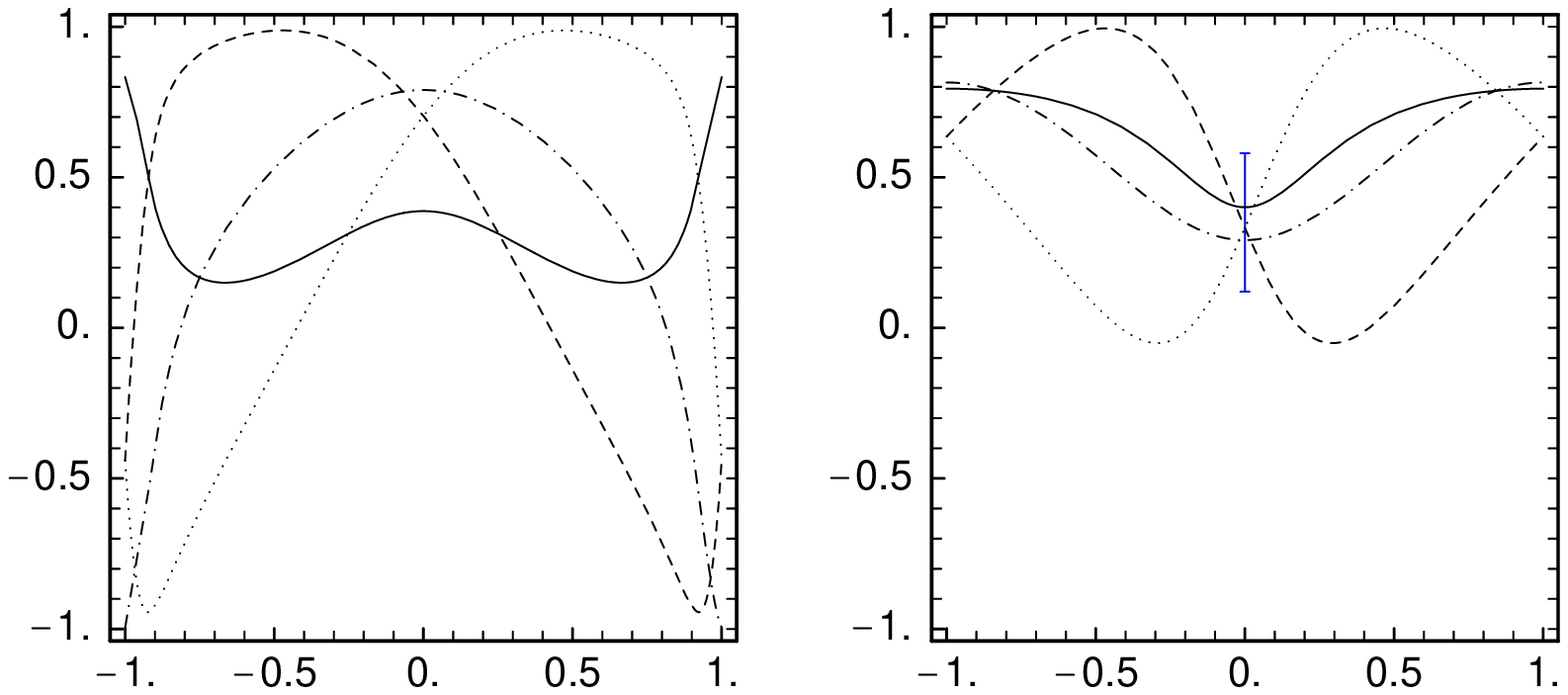,height=7cm}}}
\put(36,-1){\mbox{ $\phi_1~[\pi]$ }}
\put(109,-1){\mbox{ $\varphi_{\ti\tau}~[\pi]$}}
\put(-2,37){\mbox{ $\P_{\!\tau}^{}$ }}
\put(72,37){\mbox{ $\P_{\!\tau}^{}$ }}
\put(-2,64){\mbox{ \bf a) }}
\put(72,64){\mbox{ \bf b) }}
\end{picture} }
\end{center}
\caption{Average polarization of the tau lepton coming from
$\ti\tau_1\to\tau\nt_1$ decays for $\theta_{\ti\tau}=130^\circ$
and $\tan\beta=10$: in a) as a function of $\phi_1$ for
$M_2=380$~GeV and $|\mu|=125$~GeV; in b) as a function of
$\varphi_{\ti\tau}$ for $|\mu|=125$~GeV and $M_2$ adjusted such
that $\mnt{1}=100$~GeV. The full, dashed, dotted, and dash-dotted
lines are for $\phstau$ ($\phi_1$) $= 0,\pif,-\pif,\pi$ in a (b).
The error on $\P_{\!\tau}^{}$ indicated by the vertical bar in b)
has been estimated as described in the text.
\label{fig:Ptau_phases}}
\end{figure}

    We next choose specific values of $M_2$ and $|\mu|$ to discuss
the phase dependences in more detail. \Fig{Ptau_phases}a shows
$\P_{\!\tau}^{}$ as a function of $\phi_1$, for $M_2=380$~GeV,
$|\mu|=125$~GeV and $\phstau=0$, $\frac{\pi}{2}$, $-\frac{\pi}{2}$
and $\pi$. Since for fixed $M_2$ and $|\mu|$ the $\nt_1$ mass
changes with $\phi_1$, we show in addition in \fig{Ptau_phases}b
$\P_{\!\tau}^{}$ as a function of $\phstau$ for various values of
$\phi_1$, with $|\mu|=125$~GeV and $M_2$ adjusted such that
$\mnt{1}=100$~GeV.
   $\P_{\!\tau}^{}$ varies over a large range depending on $\phi_1$ and
$\phstau$; if the neutralino mass parameters, $\tan\b$ and
$\t_{\ti\tau}$ are known, $\P_{\!\tau}^{}$ can hence be used as a
sensitive probe of these phases (although additional information
will be necessary to resolve ambiguities and actually determine
the various phases).
    At a LC, the parameters of the neutralino/chargino sector
and also sfermion masses and mixing angles can be determined very
precisely, exploiting tunable beam energy and beam polarization
\cite{Aguilar-Saavedra:2001rg}. The actual precision depends of
course on the specific scenario.
    To illustrate the influence of uncertainties in the knowledge of
the model parameters, we take the case of $M_2=380$~GeV,
$|\mu|=125$~GeV and vanishing phases as reference point and assume
that the following precisions can be achieved:\footnote{We make a
somewhat conservative estimate because a full simulation of such a
scenario is not available.} $\d M_1=\d M_2=\d\mu=0.5\%$,
$\d\tan\b=1$, $\d\t_{\ti\tau}=3.4^\circ$, and
$\d\phi_1=\d\phi_\mu=0.1$. Varying the parameters within this
range around the reference point leads to
$\P_{\!\tau}=0.39^{+0.19}_{-0.27}$ at $\phstau=0$, which is
indicated as an error bar in \fig{Ptau_phases}b. (The 3--5\%
uncertainty in the measurement of $\P_{\!\tau}$ is comparatively
negligible). We conclude that in our particular scenario, if no
phase has been observed in the neutralino/chargino sector, a
measurement of $\P_{\!\tau}^{}$ would be sensitive to
$|\phstau|\gsim 0.3\pi$. If $\tan\b$ can be measured to
$\tan\b=10\pm 0.1$, this improves to $\d\P_{\!\tau}\simeq 0.1$ and
$|\phstau|\gsim 0.2\pi$. According to \eq{xrel}, a measurement of
a non-zero $\phstau+\phi_\mu$ implies a lower limit on $|A_\tau|$;
in our example where $\phi_\mu=0$,
$|A_\tau|>735$~GeV (1~TeV) for $|\phstau| > 0.2\pi$ ($0.3\pi$).
Increasing the precision in $\d M_1$, $\d M_1$ and $\d |\mu|$ from 0.5\%
to 0.1\% barely improves these limits.


\begin{figure}[p]
\begin{center} {\setlength{\unitlength}{1mm}
\begin{picture}(148,72)
\put(0,0){\mbox{\epsfig{figure=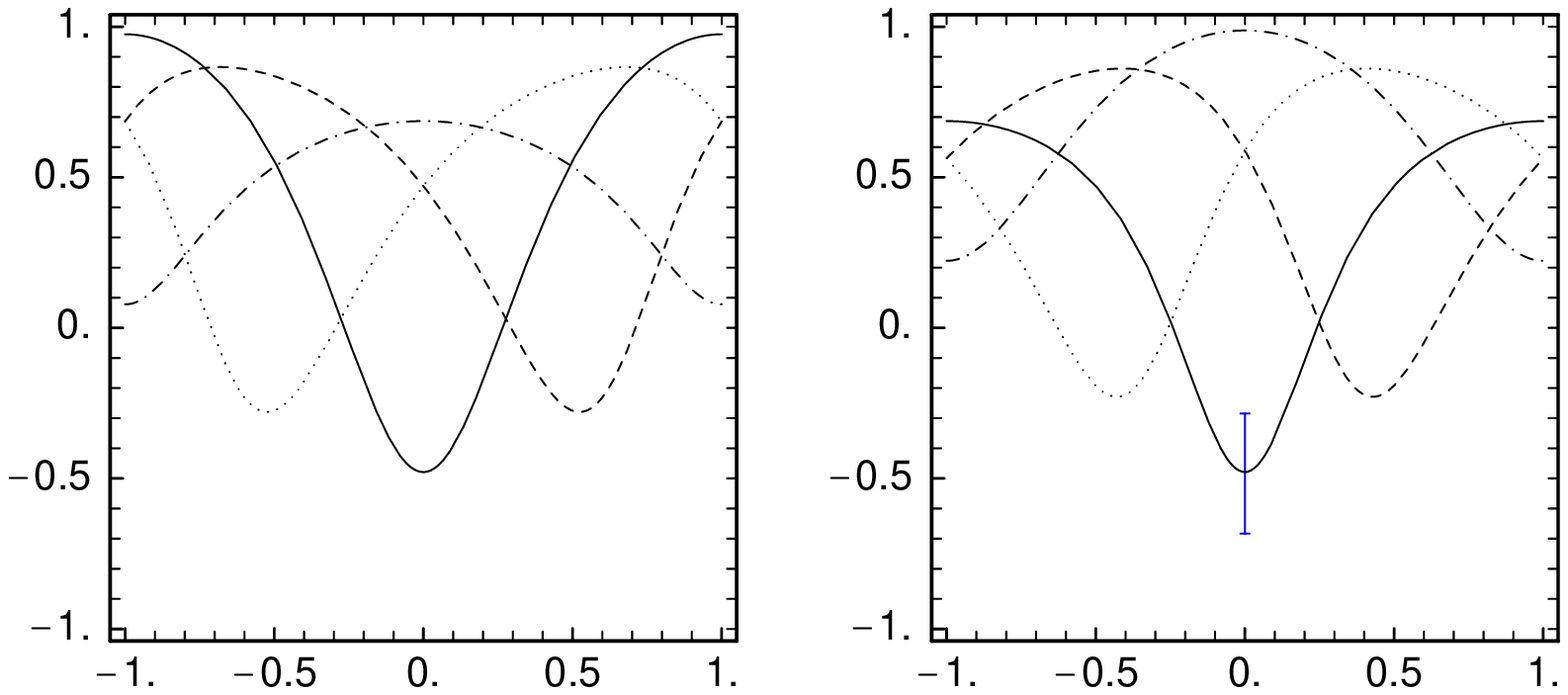,height=7cm}}}
\put(36,-1){\mbox{ $\phi_1~[\pi]$ }}
\put(108,-1){\mbox{ $\varphi_{\ti t}~[\pi]$}}
\put(-2,37){\mbox{ $\P_{\!t}^{}$ }}
\put(72,37){\mbox{ $\P_{\!t}^{}$ }}
\put(-2,64){\mbox{ \bf a) }}
\put(72,64){\mbox{ \bf b) }}
\end{picture} }
\end{center}
\caption{Average polarization of the top quark coming from $\ti
t_1\to t\nt_1$ decays for $\theta_{\ti t}=130^\circ$, and
$\tan\beta=10$: in a) as a function of $\phi_1$ for $M_2=225$~GeV
and $|\mu|=200$~GeV; in b) as a function of $\varphi_{\ti t}$ for
$|\mu|=200$~GeV and $M_2$ adjusted such that $\mnt{1}=100$~GeV.
The full, dashed, dotted, and dash-dotted lines are for $\phst$
($\phi_1$) $= 0,\pif,-\pif,\pi$ in a (b). The error on
$\P_{\!t}^{}$ indicated by the vertical bar in b) has been
estimated as described in the text. \label{fig:Ptop_phases}}
\end{figure}


We perform a similar analysis for $\P_{\!t}^{}$, using
$M_2=225$~GeV and $\mu=200$~GeV as reference point. The results
are shown in \figs{Ptop_phases}a,b in analogy to
\figs{Ptau_phases}a,b. Again a high sensitivity to both $\phi_1$
and $\phst$ is observed.
    For the case of vanishing phases, we get $\P_{\!t}^{}=-0.48$.
A variation of the parameters around the reference point as above
(with $\d\t_{\ti t}=3.5^\circ$) leads to a parametric uncertainty
of $\d\P_{\!t}^{par}\simeq 0.2$. Adding the experimental
resolution $\d\P_{\!t}^{exp}\simeq 0.1$ in quadrature gives
$\P_{\!t}^{}=-0.48\pm 0.22$ at $\phst=0$, indicated as an error
bar in \fig{Ptop_phases}b. We see that in this scenario
$\P_{\!t}^{}$ would be sensitive to $|\phst|\gsim 0.15\pi$. If
$\t_{\ti t}$ can be measured to $\sim 1^\circ$ this improves to
$\d\P_{\!t}^{par}\simeq 0.1$ ($\d\P_{\!t}^{} \simeq 0.14$) and
$|\phst|\gsim0.11\pi$; if $M_1$, $M_2$, $|\mu|$ can be measured to
0.1\% and $\tan\b$ to $0.1$, $\d\P_{\!t}^{par}$ becomes negligible
with respect to the experimental resolution of $\P_{\!t}$.
   Since $\phst\simeq\phi_{A_t}^{}$, a measurement of $\P_{\!t}^{}$
can be used to derive information on $A_t$. In particular, if both mass
eigenstates are known, $A_t$ is given by
\begin{equation}
  A_t =
  \sfrac{1}{2 m_t}
  ( \mst{2}^{2} - \mst{1}^{2} )\, |\sin2\t_{\ti t}| e^{i\phst}
  + \mu^* \cot\b \,.
\end{equation}
An analogous relation with $\cot\b\to\tan\b$ holds for $A_{\tau}$,
although the precision on $A_\tau$ is in general much worse than
on $A_t$.
In this context note that $\P_{\!f}^{}$ can also be useful to resolve
the sign ambiguity in the $\cos\t_{\ti f}$ determination from cross
section measurements \cite{Bartl:1997yi} in the CP-conserving case.
This corresponds to distinguishing the cases $\phsf=0$ and $\phsf=\pi$.

    Even though we have presented results of our analysis for $\phi_\mu^{}$,
chosen in order to satisfy the EDM constraints without having to appeal
to cancellations, we have also investigated the case of a non-zero
$\phi_\mu$. We found that a non-zero $\phi_\mu$ shifts the curves in
\figs{Ptau_M2mu}--\ref{fig:Ptop_phases} but does not cause a
qualitative change of the results.


\begin{figure}[t!]
\begin{center} {\setlength{\unitlength}{1mm}
\begin{picture}(148,72)
\put(0,0){\mbox{\epsfig{figure=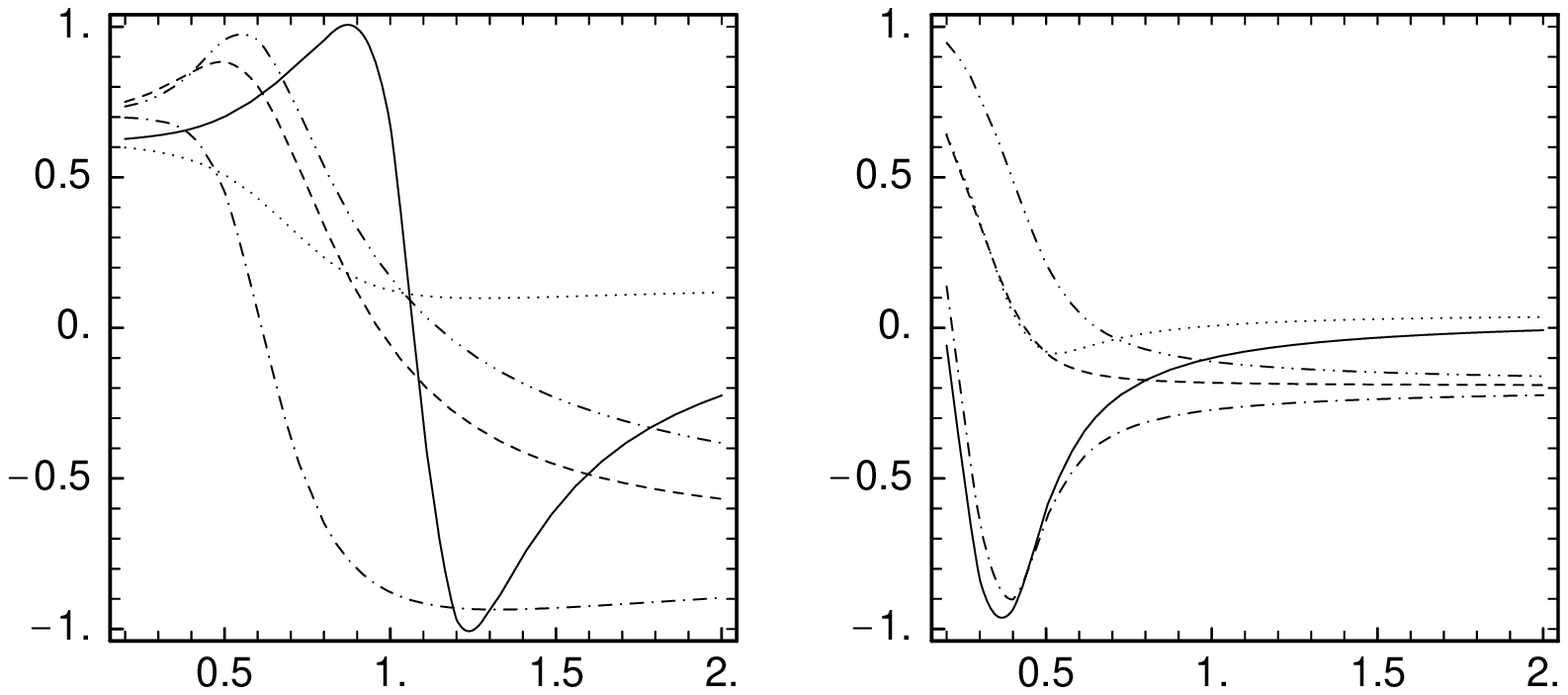,height=7cm}}}
\put(38,-1){\mbox{ $r$ }}
\put(112,-1){\mbox{ $r$ }}
\put(-2,37){\mbox{ $\P_{\!\tau}^{}$ }}
\put(72,37){\mbox{ $\P_{\!t}^{}$ }}
\put(-2,64){\mbox{ \bf a) }}
\put(72,64){\mbox{ \bf b) }}
\end{picture} }
\end{center}
\caption{$\P_{\!\tau}^{}$ (left) and $\P_{\!t}^{}$ (right)
as functions of $r=|M_1|/M_2$, for $M_2=300$~GeV and $|\mu|=150$~GeV.
The full, dashed, dotted, dash-dotted, and dash-dot-dotted
lines are for $(\phi_1,\,\varphi_{\ti \tau,\ti t})=(0,\,0)$,
$(0,\,\frac{\pi}{2})$, $(\frac{\pi}{2},0)$,
$(\frac{\pi}{2},\,\frac{\pi}{2})$ and
$(\frac{\pi}{2},\,-\frac{\pi}{2})$, respectively.
\label{fig:nogut}}
\end{figure}


    Last but not least we note that giving up the GUT relation between
$|M_1|$ and $M_2$ changes the picture completely, as the pattern
of gaugino--higgsino mixing is strongly affected
\cite{Belanger:2001am}. This is illustrated in \fig{nogut}, where
we plot $\P_{\!\tau}^{}$ and $\P_{\!t}^{}$ as functions of
$r=|M_1|/M_2$ for $M_2=300$~GeV, $|\mu|=150$~GeV and the other
parameters as in \figs{Ptau_M2mu} and \ref{fig:Ptop_M2mu}. A
detailed study of the implications of non-universal gaugino masses
will be presented elsewhere.\\

{\it To sum up,} both $\P_{\!t}^{}$ and $\P_{\!\tau}^{}$ can vary
over a large range depending on $\phi_1$ and $\varphi_{\ti
t,\ti\tau}$ (and also $\phi_\mu$, though we did not discuss this
case explicitly) and may thus be used as sensitive probes of these
phases. To this aim, however, the neutralino mass parameters,
$\tan\b$ and the sfermion mixing angles need to be known. Given
the complexity of the problem, a combined fit of all available
data seems to be the most convenient method for the extraction of
the MSSM parameters.

\section{\mbf Fermion polarization in $\sf\to f'\ch$ decays} \label{sect:Pfpr}

The sfermion interaction with charginos is ($i,j=1,2$)
\begin{eqnarray}
  \L_{f'\!\sf\ch}
  &=& g\,\bar u\,( l_{ij}^{\,\ti d}\,\PR +
                   k_{ij}^{\,\ti d}\,\PL )\,\ti\x^+_j\,\ti d_i^{}
    + g\,\bar d\,( l_{ij}^{\,\ti u}\,\PR +
                   k_{ij}^{\,\ti u}\,\PL )\,\ti\x^{+c}_j\,\ti u_i^{}
      + {\rm h.c.}
\end{eqnarray}
where $u$ ($\ti u$) stands for up-type (s)quark and (s)neutrinos,
and $d$ ($\ti d$) stands for down-type (s)quark and charged (s)leptons.
The couplings $l$ and $k$ are
\begin{align}
  l_{ij}^{\ti t} &= -V_{j1} R_{i1}^{\ti t\,*} + h_t\,V_{j2} R_{i2}^{\ti t\,*}\,, &
  l_{ij}^{\ti b} &= -U_{j1} R_{i1}^{\ti b\,*} + h_b\,U_{j2} R_{i2}^{\ti b\,*}\,,
  \label{eq:elltb} \\
  k_{ij}^{\ti t} &= h_b\,U_{j2}^* R_{i1}^{\ti t\,*} \,, &
  k_{ij}^{\ti b} &= h_t\,V_{j2}^* R_{i1}^{\ti b\,*} \,,
  \label{eq:katb}
\end{align}
for stops and sbottoms and
\begin{align}
  l_{j}^{\ti\nu}  &= -V_{j1} \,, &
  l_{ij}^{\ti\tau} &= -U_{j1} R_{i1}^{\ti\tau\,*} + h_\tau\,U_{j2}R_{i2}^{\ti\tau\,*}\,, \\
  k_{j}^{\ti\nu}  &= h_\tau\,U_{j2}^* \,, &
  k_{ij}^{\ti\tau} &= 0 \,
\end{align}
for staus and sneutrinos.

Analogous to the decay into a neutralino, eq.~\eq{Pf}, the average
polarization of the fermion coming from the $\sf_i\to f'\ch_j$
decay is given by
\begin{equation}
   {\P_{\!f}^{}}'
           = \frac{Br\,(\sf_i\to\ch_j f_R')-Br\,(\sf_i\to\ch_j f_L')}
                  {Br\,(\sf_i\to\ch_j f_R')+Br\,(\sf_i\to\ch_j f_L')}
           = \frac{|k_{ij}^{\sf}|^2-|l_{ij}^{\,\sf}|^2}
                  {|k_{ij}^{\sf}|^2+|l_{ij}^{\,\sf}|^2} \,.
\label{eq:Pfprime}
\end{equation}

\begin{figure}[t!]
\begin{center} {\setlength{\unitlength}{1mm}
\begin{picture}(70,74)
\put(0,0){\mbox{\epsfig{figure=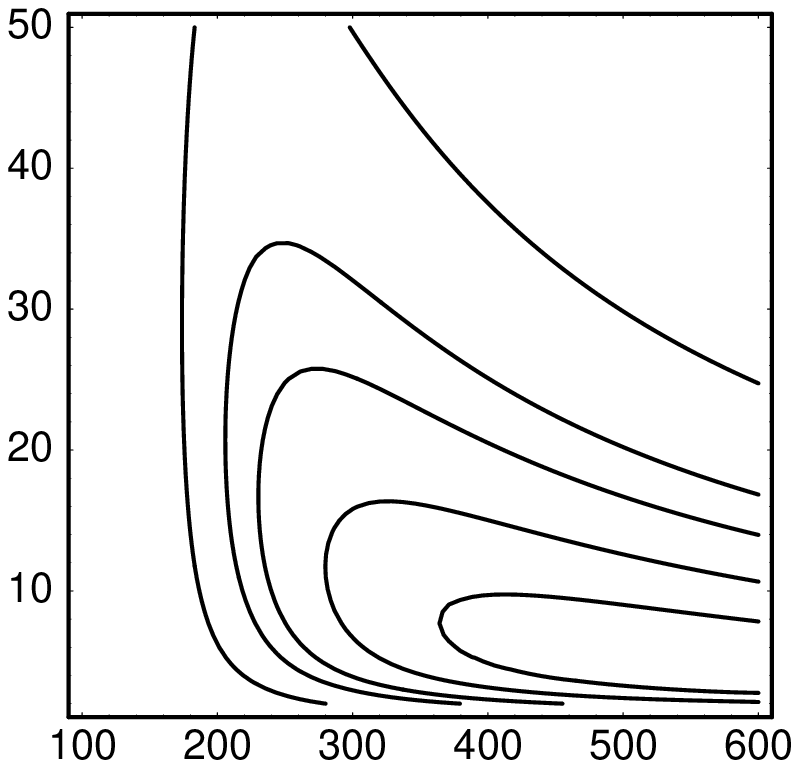,height=7cm}}}
\put(27,-3){\mbox{ $|\mu|$~[GeV] }}
\put(-2,71){\mbox{ $\tan\beta$ }}
\put(7,60){\mbox{ 0.01 }}
\put(36,60){\mbox{ 0.01 }}
\put(22,50){\mbox{ 0.03 }}
\put(24,38){\mbox{ 0.05 }}
\put(28,26.5){\mbox{ 0.1 }}
\put(34,18){\mbox{ 0.2 }}
\end{picture} }
\end{center}
\caption{Contours of constant $\Delta({\P_{\!\tau}^{}}')_2^{} =
 {\P_{\!\tau}^{}}'(\ti\nu_\tau\to\tau\ti\x^+_2,\phi_\mu=\frac{\pi}{2}) -
 {\P_{\!\tau}^{}}'(\ti\nu_\tau\to\tau\ti\x^+_2,\phi_\mu=0)$
 in the $(\tan\beta,\,|\mu|)$ plane for $M_2=150$~GeV.
\label{fig:Ptauprime}}
\end{figure}

\noindent
Since only top and tau polarizations are measurable, we
only discuss $\ti b\to t\ti\x^-$ and $\ti\nu_\tau\to\tau\ti\x^+$
decays. The latter case is especially simple because
${\P_{\!\tau}^{}}'$ depends only on the parameters of the chargino
sector:
\begin{equation}
   ({\P_{\!\tau}^{}}')_j^{} = {\P_{\!\tau}^{}}'(\ti\nu_\tau\to\tau\ti\x^+_j)
            = \frac{|h_\tau U_{j2}|^2-|V_{j1}|^2}
                   {|h_\tau U_{j2}|^2+|V_{j1}|^2} \,.
\end{equation}
A measurement of ${\P_{\!\tau}^{}}'$ may hence be useful to
supplement the chargino parameter determination. However, the
dependence of ${\P_{\!\tau}^{}}'(\ti\nu_\tau\to\tau\ti\x^+_1)$ on
$\phi_\mu^{}$ turns out to be very small, the effects being in
general well below 1\% (i.e. $\Delta\P < 0.01$). Only for the
decay into the heavier chargino, the effect of a non-zero
phase\footnote{We remind the reader that unless huge cancellations
are invoked, $\phi_\mu^{}$ is severely restricted by the
non-observation of the eEDM.} may be sizeable. As an example,
\fig{Ptauprime} shows the differences in
$({\P_{\!\tau}^{}}')_2^{}$ between $\phi_\mu=0$ and
$\phi_\mu=\pi/2$ in the $(\tan\beta,\,|\mu|)$ plane for
$M_2=150$~GeV. $\Delta({\P_{\!\tau}^{}}')_2^{} =
 {\P_{\!\tau}^{}}'(\ti\nu_\tau\to\tau\ti\x^+_2,\phi_\mu=\frac{\pi}{2}) -
 {\P_{\!\tau}^{}}'(\ti\nu_\tau\to\tau\ti\x^+_2,\phi_\mu=0)$ can go up
to $\sim 0.25$. However, it requires quite heavy sneutrinos for
this decay to be kinematically allowed. Moreover, the measurement
of $({\P_{\!\tau}^{}}')_2^{}$ will be diluted by
$\ti\nu_\tau\to\tau\ti\x^+_1$
decays. \\


Let us now turn to top polarization in $\ti b\to t\ti\x^-$ decays.
For $\ti b_1^{}$ decays, we have
\begin{eqnarray}
  |k_{1j}^{\ti b}|^{2} - |l_{1j}^{\ti b}|^{2}
  &=&  h_t^2 \cos^2\theta\, | V_{j2}^*\,e^{-i\varphi} |^2
       - | h_b U_{j2}^{}\sin\theta - U_{j1}\cos\theta\,e^{-i\varphi} |^{2}
  \nn\\
  &=&  ( h_t^2 |V_{j2}^{}|^2 - |U_{j1}^{}|^2 ) \cos^2\theta
       - h_b^2 |U_{j2}^{}|^2 \sin^2\theta \nn\\
  & & +\, h_b \sin 2\theta \,
      ( \Re[ U_{j1}^{} U_{j2}^{*} ] \cos\varphi
       +\Im[ U_{j1}^{} U_{j2}^{*} ] \sin\varphi ) \,.
\label{eq:sbot1}
\end{eqnarray}
For $\ti b_2^{}$ decays, the corresponding expression is given by
the RHS of \eq{sbot1} with $\cos^2\theta$, $\sin^2\theta$
interchanged, and a change in sign of the term $\propto\sin
2\theta$. We see that the phase dependence of ${\P_{\!t}^{}}'$ is
proportional to $h_b \sin 2\theta_{\ti b}$ and the amount of
gaugino--higgsino mixing of the charginos; it will therefore be
largest for $|M_2|\sim|\mu|$, $\theta_{\ti b}=3\pi/4$ and large
$\tan\beta$. Again, there is a non-zero effect even if there is
just one phase in either the sbottom or chargino sector.
    Note, however, that the only CP phase in the chargino sector
is $\phi_\mu$, which also enters the sfermion mass matrices.
Complex $U$ and $V$ hence imply $\phsb\not=0$. More precisely,
$a_b = |a_b|e^{i\varphi_{\ti b}}\sim \mu^*\tan\beta
     = |\mu|\tan\beta\,e^{-i\phi_\mu}$
for medium and large $\tan\beta$, and thus $\phsb= -\phi_\mu^{}$ unless
$|A_b|\gsim |\mu|\tan\b$; see eq.~\eq{xrel} and the related discussion.
For the sake of a general discussion of the phase dependence
of ${\P_{\!t}^{}}'$ (and since $A_b$ is still a free parameter),
we nevertheless use $\phi_\mu^{}$ and $\phsb$ as independent
input parameters.

\begin{figure}[p]
\begin{center} {\setlength{\unitlength}{1mm}
\begin{picture}(148,72)
\put(0,0){\mbox{\epsfig{figure=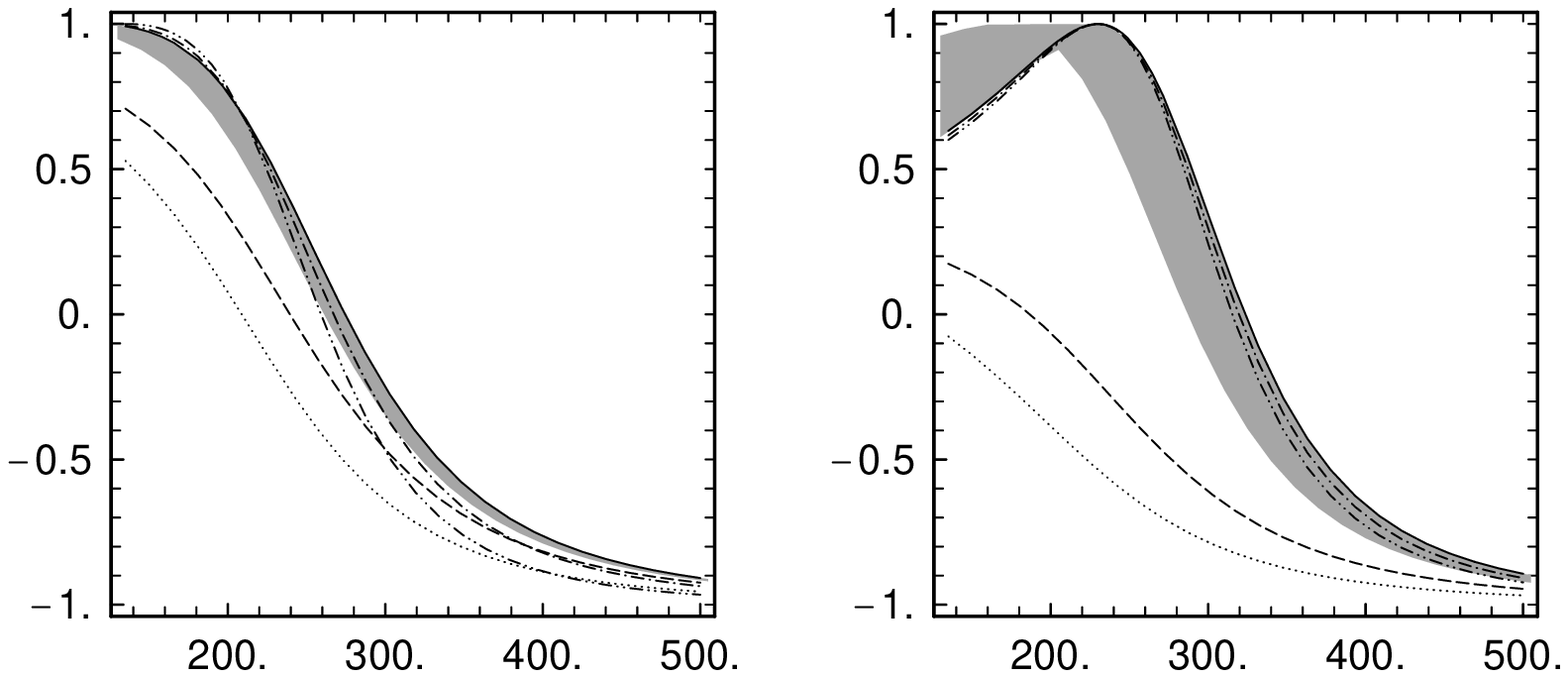,height=7cm}}}
\put(32,-1){\mbox{ $|\mu|$~[GeV] }}
\put(104,-1){\mbox{ $|\mu|$~[GeV] }}
\put(-2,37){\mbox{ ${\P_{\!t}^{}}'$ }}
\put(72,37){\mbox{ ${\P_{\!t}^{}}'$ }}
\put(-2,64){\mbox{ \bf a) }}
\put(72,64){\mbox{ \bf b) }}
\put(30,67){\mbox{ $\tan\beta=10$ }}
\put(102,67){\mbox{ $\tan\beta=30$ }}
\end{picture} }
\end{center}
\caption{Average polarization of the top quark coming from $\ti
b_1^{}\to t\ti\x^-_1$ decays as a function of $|\mu|$ for
$\theta_{\ti b}=140^\circ$, $M_2 = 225$~GeV, $\tan\beta=10$ in a)
and $\tan\beta=30$ in b). The full, dashed, dotted, dash-dotted
and dash-dot-dotted lines are for $(\phi_\mu,\,\varphi_{\ti
b})=(0,\,0)$, $(0,\,\frac{\pi}{2})$,
$(\frac{\pi}{2},\frac{\pi}{2})$,
$(-\frac{\pi}{2},\,\frac{\pi}{2})$, and $(-\pi,\pi)$ respectively.
The grey bands show the range of ${\P_{\!t}^{}}'$ due to varying
$m_b$ within 2.5--4.5 GeV for the case $\phi_\mu=0$.
\label{fig:Ptoppr_M2mu}}
\end{figure}

    \Fig{Ptoppr_M2mu} shows the average top polarization in
$\ti b_1^{}\to t\ti\x^-_1$ decays as a function of $|\mu|$ for
$M_2 = 225$~GeV, $\tan\beta=10$ and 30, and various combinations
of $\phi_\mu^{}$ and $\phsb$. Here we have fixed $\theta_{\ti
b}=140^\circ$, since from renormalization-group running one
expects $\msb{L}<\msb{R}$. As in the previous section, we find
large effects from CP-violating phases if the $\ch$ has a sizeable
higgsino component; as expected, these effects are enhanced for
large $\tan\b$. The results stay the same if both $\phi_\mu^{}$
and $\phsb$ change their signs. Moreover,
${\P_{\!t}^{}}'(\phi_\mu=0,\,\varphi_{\ti b}=\frac{\pi}{2}) \sim
{\P_{\!t}^{}}'(\phi_\mu=\frac{\pi}{2},\,\varphi_{\ti b}=0)$. If
$\phi_\mu^{}$ and $\phsb$ have the same sign, the difference in
${\P_{\!t}^{}}'$ from the case of vanishing phases is larger than
if they have opposite signs. In particular, we find
${\P_{\!t}^{}}'(\phi_\mu=-\varphi_{\ti b})\sim
{\P_{\!t}^{}}'(\phi_\mu=\varphi_{\ti b}=0)$ over large regions of
the parameter space. With an experimental resolution of the top
polarization of about 10\% this implies that in many cases
$\varphi_{\ti b}\sim -\phi_\mu$ cannot be distinguished from
$\varphi_{\ti b}=\phi_\mu=0$ by measurement of ${\P_{\!t}^{}}'$.
Furthermore, the value of ${\P_{\!t}^{}}'$ is quite sensitive to
the running $b$ quark mass, which enters the bottom Yukawa
coupling of eq.~\eq{yuk} and is subject to possibly large SUSY
loop corrections. For the lines in \fig{Ptoppr_M2mu} we have used
$m_b=4.5$~GeV. The grey bands show the range of ${\P_{\!t}^{}}'$
when $m_b$ is varied between 2.5 and 4.5 GeV. As can be seen, the
uncertainty in $m_b$ --- more precisely in $h_b$ --- tends to wash
out small effects of CP-violating phases, specially in the case of
large $\tan\b$.


\begin{figure}[p]
\begin{center} {\setlength{\unitlength}{1mm}
\begin{picture}(148,72)
\put(5,3.3){\mbox{\epsfig{figure=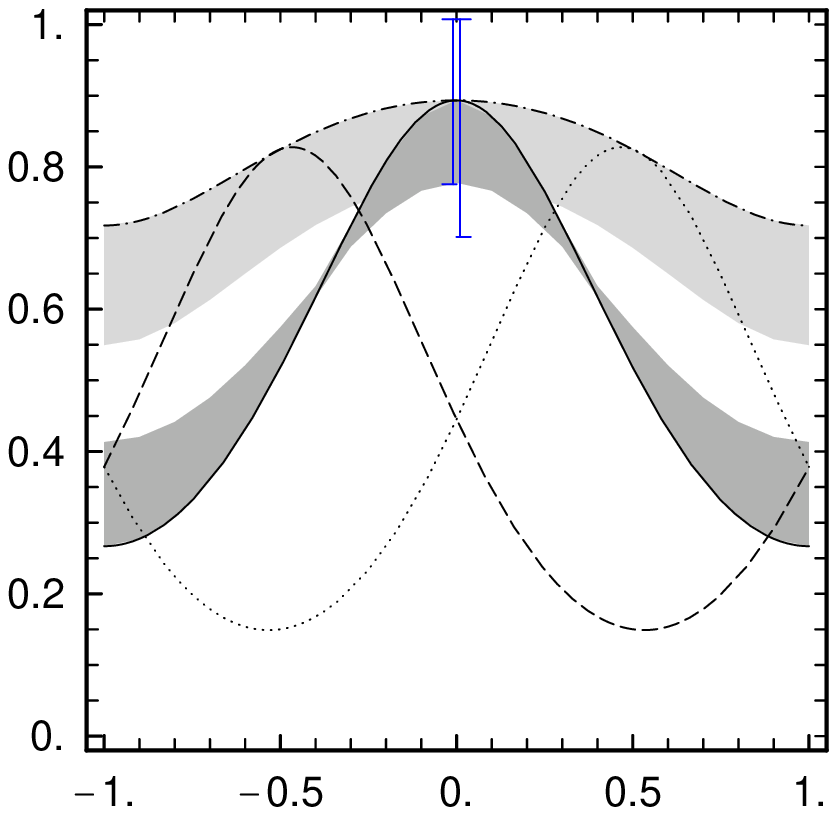,height=62.5mm}}}
\put(81,2){\mbox{\epsfig{figure=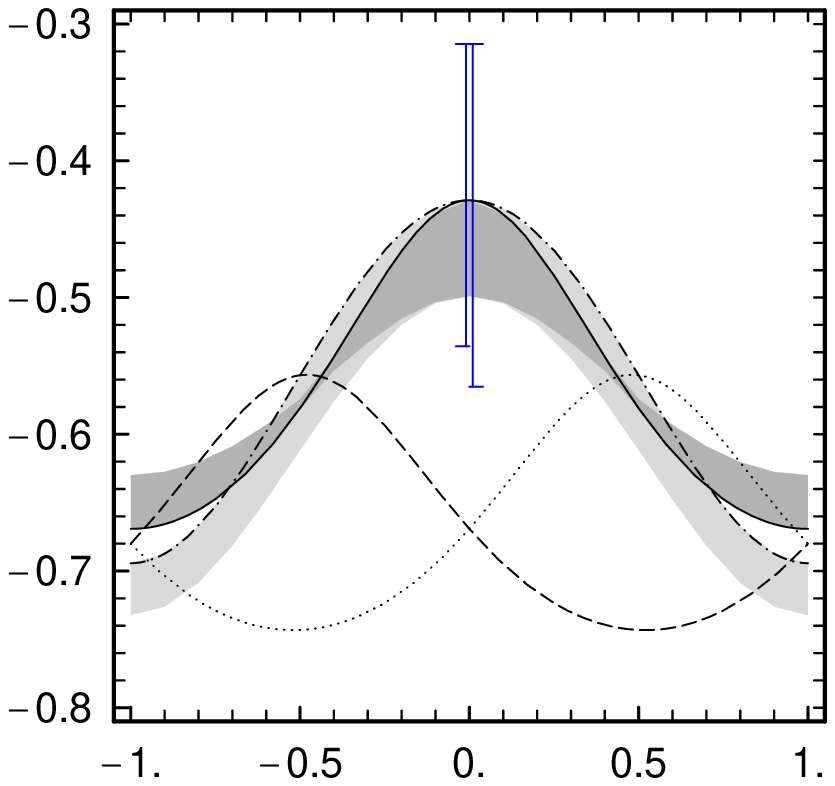,height=65mm}}}
\put(33,-1){\mbox{ $\varphi_{\ti b}~[\pi]$ }}
\put(112,-1){\mbox{ $\varphi_{\ti b}~[\pi]$ }}
\put(-4,37){\mbox{ ${\P_{\!t}^{}}'$ }}
\put(73,37){\mbox{ ${\P_{\!t}^{}}'$ }}
\put(-4,63){\mbox{ \bf a) }}
\put(73,63){\mbox{ \bf b) }}
\put(14,68){\mbox{ $|\mu|=200$, $\mch{1}=155$~GeV }}
\put(92,68){\mbox{ $|\mu|=350$, $\mch{1}=204$~GeV }}
\end{picture} }
\end{center}
\caption{Average polarization of the top quark coming from $\ti
b_1^{}\to t\ti\x^-_1$ decays as a function of $\varphi_{\ti b}$,
for $\theta_{\ti b}=140^\circ$, $\tan\beta=10$, $|\mu|=200$ and
$\mch{1}=155$~GeV in a) and $|\mu|=350$ and $\mch{1}=204$~GeV in
b). The full, dashed and dotted lines are for $\phi_\mu = 0$,
$\frac{\pi}{2}$ and $-\frac{\pi}{2}$, respectively, while for the
dash-dotted lines $\phi_\mu = -\varphi_{\ti b}$. The grey bands
show the range of ${\P_{\!t}^{}}'$ due to varying $m_b$ within
2.5--4.5 GeV for the cases $\phi_\mu=0$ and $\phi_\mu =
-\varphi_{\ti b}$. The error bars show the estimated errors on
${\P_{\!t}^{}}'$ as described in the text. \label{fig:Ptoppr_phi}}
\end{figure}


    In order to see what information can be extracted from a ${\P_{\!t}^{}}'$
measurement, we pick two values of $|\mu|$ from
\fig{Ptoppr_M2mu}a; namely $|\mu|=200$~GeV and $|\mu|=350$~GeV,
and show the phase dependences at these points in
\fig{Ptoppr_phi}.
    \Fig{Ptoppr_phi}a shows ${\P_{\!t}^{}}'$ as a function of $\phsb$,
for $|\mu|=200$~GeV, $\tan\beta=10$, $\theta_{\ti b}=140^\circ$,
and various values of $\phi_\mu$. $M_2$ is chosen such that
$\mch{1}=155$~GeV (i.e.\ $M_2=225$~GeV for $\phi_\mu=0$). The
range obtained by varying $m_b$ within 2.5--4.5~GeV is shown as
grey bands for two of the curves, for $\phi_\mu=0$ and
$\phi_\mu=-\phsb$. We estimate the effect of an imperfect
knowledge of the model parameters in the same way as in the
previous section. For $M_2=225\pm 1.125$~GeV, $|\mu|=200\pm
1$~GeV, $\tan\b=10\pm 1$, $\t_{\ti b}=140\pm 3.4^\circ$ and
$\phi_\mu=0\pm 0.1$, we get ${\P_{\!t}^{}}'=0.89 \pm 0.06$ at
$\phsb=0$. Varying in addition $m_b=2.5$--4.5 GeV gives
${\P_{\!t}^{}}'=0.89^{+0.06}_{-0.16}$. Adding a 10\% measurement
error on ${\P_{\!t}^{}}'$ in quadrature, we end up with
${\d\P_{\!t}^{}}'=0.12$ ($0.19$) without (with) the $m_b$ effect.
These are shown as error bars in \fig{Ptoppr_phi}a.
We see that the case of $\phsb=-\phi_\mu^{}$ cannot be
distinguished from $\phsb=\phi_\mu^{}=0$ in this scenario.
However, ${\P_{\!t}^{}}'$ turns out to be quite a sensitive probe
of $\delta_\phi^{} = \phsb+\phi_\mu^{}$, i.e.\ the deviation from
the `natural' alignment $\phsb=-\phi_\mu^{}$. In the example of
\fig{Ptoppr_phi}a, $|\delta_\phi^{}|\gsim 0.24\pi$ $(0.31\pi)$ can
be resolved if $h_b$ is (not) known precisely, quite independently
of $\phi_\mu^{}$. Observing such a $\delta_\phi^{}$ also implies a
bound on $|A_b|$ of $|A_b| > 1363$ $(1678)$~GeV. If the precision
on $M_2$ and $|\mu|$ is 0.1\% and $\tan\b=10\pm 0.1$, we get
$({\d\P_{\!t}^{}}')^{par}= 0.03$ at $\phsb=0$, so that the error
is dominated by the experimental uncertainty. However, the
resultant improvement in the sensitivity is limited to
$|\delta_\phi^{}|\gsim 0.22\pi$ and $|A_b| > 1294$~GeV.

    \Fig{Ptoppr_phi}b shows ${\P_{\!t}^{}}'$ as a function of
$\phsb$, for $|\mu|=350$~GeV, $\mch{1}=204$~GeV, and the other
parameters as above ($M_2=225$~GeV at $\phi_\mu=0$). The effect of
an uncertainty in $h_b$ is again shown as grey bands for
$\phi_\mu=0$ and $\phi_\mu=-\phsb$. Estimating the parametric
uncertainty in the same way as above, but with $|\mu|=350\pm
1.75$~GeV, we get ${\P_{\!t}^{}}'=-0.43^{+0.06}_{-0.04}$ at
$\phsb=0$. Varying in addition $m_b=2.5$--4.5 GeV gives
${\P_{\!t}^{}}'=-0.43^{+0.06}_{-0.09}$. Adding a 10\% measurement
error on ${\P_{\!t}^{}}'$ in quadrature, we end up with
${\d\P_{\!t}^{}}'=0.11$ ($0.14$) without (with) the effect of
$m_b$, shown as error bars in \fig{Ptoppr_phi}b.
    In a three-dimensional plot, ${\P_{\!t}^{}}'$ has a bell-like shape
in the $\phi_\mu$--$\phsb$ plane, with contours of constant
${\P_{\!t}^{}}'$ being ellipses in this plane. If $\phi_\mu^{}$ is
not known, a measurement of ${\P_{\!t}^{}}'$ may therefore be
useful to put limits on $\phi_\mu$ and $\phsb$, but not on
$\delta_\phi^{}$, which restricts $|A_b|$. In our case study, we
have assumed that $\phi_\mu^{}=0$ is known. In this case, a
measurement of ${\P_{\!t}^{}}'=-0.43$, for instance, would
restrict $|\phsb| \lsim 0.38\pi$ at $1\sigma$, while a measurement
of ${\P_{\!t}^{}}'=-0.56$ would disfavour $|\phsb|\gsim 0.9\pi$ as
well as $|\phsb|\lsim 0.13\pi$. The latter would also allow a
constraint on $\delta_\phi^{}$. As mentioned above, a lower limit
on $\delta_\phi^{}$ implies a lower limit on $|A_b|$. An upper
limit on $\d_\phi{}$ can be used to set an upper limit on $|A_b|$
as a function of $\phi_{A_{b}}$: $|A_b| <
|\frac{\sin(\phsb+\phi_\mu^{}) }{\sin(\phsb-\phi_{A_{b}})}|
         \times |\mu| \tan\b$.
Note, however, that this becomes unbounded for $\phi_{A_{b}} \to \phsb$.

    We have also investigated the case of large $\tan\b$ ($\tan\b=30$).
It reveals a $\phsb$ dependence similar to that of
\fig{Ptoppr_phi}a, with almost no dependence on $\phi_\mu^{}$ and
the $\d_\phi^{}$ dependence accordingly more pronounced. We
encounter, however, a large parametric uncertainty, which
practically washes out the sensitivity to $\d_\phi{}$.\\

{\it To sum up,} tau polarization in $\ti\nu_\tau\to\tau\ti\x^+$
decays depends only little on $\phi_\mu^{}$. ${\P_{\!\tau}^{}}'$
is hence not a promising quantity to study CP phases, but may be
useful for (consistency) tests of the gaugino--higgsino mixing.
Top polarization in $\ti b\to t\ti\x^-$ decays, on the other hand,
can be useful to probe $\phi_\mu^{}$, $\phsb$ and/or
$\d_\phi^{}=\phi_\mu^{}+\phsb$ in some regions of the parameter
space. The measurement of ${\P_{\!t}^{}}'$, revealing phases or
being consistent with vanishing phases, may also constrain
$|A_b|$.

\section{Conclusions}

We have discussed the influence of CP-violating phases on the
fermion polarization in sfermion decays to neutralinos or
charginos, $\ti f_i\to f\,\nt_n$ and $\ti f_i\to f'\,\ch_j$
($i,j=1,2$; $n=1,...,4$; $f\!,f'=t,\tau$). This polarization is
considered as a useful tool for the MSSM parameter determination
\cite{Nojiri:1994it,Nojiri:1996fp,Boos:2003vf}.

In decays into charginos, the polarization depends on the phase of
$\mu$. Since this dependence is weak in the case of
$\ti\nu_\tau\to\tau\ti\x^+$, ${\P_{\!\tau}^{}}'$ does not provide
a promising probe of CP phases (on the other hand, exactly this
feature can make ${\P_{\!\tau}^{}}'$ useful for consistency tests
of gaugino--higgsino mixing). In $\ti b\to t\x^-$ decays, the
dependence on $\phi_\mu^{}$ can be rather large; in addition, also
the phase of the sbottom-mixing matrix plays a role. If
$|A_b|<|\mu|\tan\b$, $\phsb\simeq -\phi_\mu^{}$. We found that
this case can be difficult to distinguish from the CP-conserving
case by measuring ${\P_{\!t}}'$. If, however, a deviation from
$\phsb+\phi_\mu^{}=0$ is observed, these phases can be constrained
and also limits on $A_b$ can be derived.

The decays $\ti t\to t\nt$ and $\ti\tau\to\tau\nt$ provide a more effective
probe of CP violation because an additional phase, the phase of $M_1$,
contributes. We found that CP phases can have a significant effect on the
top and tau polarizations, especially if the involved neutralino has a sizeable
higgsino component. If the parameters of the neutralino sector can be
measured precisely, e.g.\ in $e^+e^-$ annihilation with polarized
beams, $\P_t^{}$ and $\P_\tau^{}$ can be useful 
for the determination of CP phases. In particular, since
$\phst\simeq\phi_{A_t}^{}$ unless $|\mu|$ is very large,
a measurement of $\P_{\!t}^{}$ can give information on $A_t$.

In this respect it is important to note that (for fixed masses)
the sfermion production cross sections do not depend on CP phases.
In the sfermion sector, these can be manifest in branching ratios
as discussed in
\cite{Bartl:2002bh,Bartl:2002uy,Bartl:2003he,Bartl:2003pd},
polarization of the decay fermions as discussed in this paper, and
CP-odd asymmetries. Branching ratios are in general rather
difficult to measure with high precision. The information that can
be gained from branching ratios is also limited if one decay
channel dominates, e.g. $\ti\tau_1\to\tau\nt_1$
in case of a light stau. 
This makes the polarization of the decay fermions a very
interesting possibility to explore CP phases.
Last but not least we note that the computations in this paper,
leading to effects of a few percent, have been performed at tree
level. The influence of radiative corrections
\cite{Majerotto:2002iu} can be of comparable size and will
therefore have to be taken into account for precision analyses.

A measurement of the CP phases in the sfermion/$\ch, \nt$ sector
will also complement CP studies of the Higgs sector
\cite{Godbole:2004xe}, since in the MSSM Higgs-sector CP violation
is generated through quantum corrections
\cite{Pilaftsis:1998pe,Demir:1999hj,Choi:2000wz,Carena:2000yi}.
Last but not least we emphazise that, since the effects can be
large, the possibility of CP violation should be taken into
account in precision SUSY parameter analyses, especially in a
general analysis project as envisaged in \cite{spa}.

\section*{Acknowledgements}
T.G. acknowledges the financial support of the Institut f\"ur
Hochenergiephysik in Vienna and the CERN Theory Division.  R.G.
wishes to thank the CERN Theory Division and LAPTH for hospitality
and financial support, where this work was completed. S.K. thanks the 
Centre for High Energy Physics in Bangalore for hospitality and 
financial support.  RG wishes to acknowledge the partial support of the 
Department of Science and Technology, India, under project number 
SP/S2/K-01/2000-II.

\begin{appendix}
\section{Standard Model constants}

The SM constants used in the numerical analysis are:
\begin{eqnarray}
  m_t &=& 175~{\rm GeV} \nn\\
  m_b &=& 4.5~{\rm GeV}\nn\\
  m_\tau &=& 1.77~{\rm GeV}\nn\\
  m_Z &=& 91.2~{\rm GeV}\nn\\
  m_W &=& 80.03~{\rm GeV}\nn\\
  \sin^2\t_W &=& 0.23 \nn\\
  \a\, (m_Z) &=& 1/129
\end{eqnarray}

\end{appendix}



\begin{thebibliography}{99}


\bibitem{kaon}
J.H. Christenson, J.W. Cronin, V.L. Fitch and R. Turlay,
    Phys. Rev. Lett. {\bf 13}, 138 (1964).

\bibitem{bellebabar}
BELLE Collab., A. Abashian et al.,
    Phys. Rev. Lett. {\bf 86}, 2509 (2001), hep-ex/0102018;
BABAR Collab., B. Aubert et al.,
    Phys. Rev. Lett. {\bf 86}, 2515 (2001), hep--ex/0102030.

\bibitem{Sakharov:dj}
A.~D.~Sakharov,
    Pisma Zh.\ Eksp.\ Teor.\ Fiz.\  {\bf 5} (1967) 32, JETP Lett. 6, 24 (1967).

\bibitem{Dolgov:2002kw}
For a recent summary, see A.~D.~Dolgov,
    hep-ph/0211260.

\bibitem{Bennett:2003bz}
C.~L.~Bennett {\it et al.},
    Astrophys.\ J.\ Suppl.\  {\bf 148}, 1 (2003)
    [astro-ph/0302207].

\bibitem{Dine:2003ax}
For a review, see e.g.\
    M.~Dine and A.~Kusenko,
    Rev.\ Mod.\ Phys.\  {\bf 76}, 1 (2004) [hep-ph/0303065].


\bibitem{Haber:1984rc} See, e.g.\
H.~E.~Haber and G.~L.~Kane,
    Phys.\ Rept.\  {\bf 117} (1985) 75.


\bibitem{Ibrahim:1997nc}
T.~Ibrahim and P.~Nath,
    Phys.\ Lett.\ B {\bf 418}, 98 (1998) [hep-ph/9707409];
    Phys.\ Rev.\ D {\bf 57}, 478 (1998),
    Erratum-ibid.\ D {\bf 58}, 019901 (1998)
    [hep-ph/9708456];
    Phys.\ Rev.\ D {\bf 58}, 111301 (1998),
    Erratum-ibid.\ D {\bf 60}, 099902 (1999)
    [hep-ph/9807501];
    Phys.\ Rev.\ D {\bf 61}, 093004 (2000) [hep-ph/9910553].

\bibitem{Brhlik:1998zn}
M.~Brhlik, G.~J.~Good and G.~L.~Kane,
    Phys.\ Rev.\ D {\bf 59}, 115004 (1999) [hep-ph/9810457].

\bibitem{Bartl:1999bc}
A.~Bartl, T.~Gajdosik, W.~Porod, P.~Stockinger and H.~Stremnitzer,
    Phys.\ Rev.\ D {\bf 60}, 073003 (1999) [hep-ph/9903402].

\bibitem{Falk:1998pu}
T.~Falk and K.~A.~Olive,
    Phys.\ Lett.\ B {\bf 439}, 71 (1998) [hep-ph/9806236].

\bibitem{Falk:1999tm}
T.~Falk, K.~A.~Olive, M.~Pospelov and R.~Roiban,
    Nucl.\ Phys.\ B {\bf 560}, 3 (1999) [hep-ph/9904393].


\bibitem{Pilaftsis:1998pe}
A.~Pilaftsis,
    Phys.\ Rev.\ D {\bf 58} (1998) 096010 [hep-ph/9803297];
    Phys.\ Lett.\ B {\bf 435} (1998) 88 [hep-ph/9805373];
A.~Pilaftsis and C.~E.~Wagner,
    Nucl.\ Phys.\ B {\bf 553}, 3 (1999) [hep-ph/9902371].

\bibitem{Demir:1999hj}
D.~A.~Demir,
    Phys.\ Rev.\ D {\bf 60} (1999) 055006 [hep-ph/9901389].

\bibitem{Choi:2000wz}
S.~Y.~Choi, M.~Drees and J.~S.~Lee,
    Phys.\ Lett.\ B {\bf 481}, 57 (2000) [hep-ph/0002287].

\bibitem{Carena:2000yi}
M.~Carena, J.~R.~Ellis, A.~Pilaftsis and C.~E.~M.~Wagner,
    Nucl.\ Phys.\ B {\bf 586}, 92 (2000) [hep-ph/0003180].

\bibitem{Chang:1998uc}
D.~Chang, W.~Y.~Keung and A.~Pilaftsis,
    Phys.\ Rev.\ Lett.\  {\bf 82}, 900 (1999) [Erratum-ibid.\  {\bf
    83}, 3972 (1999)] [arXiv:hep-ph/9811202].

\bibitem{Pilaftsis:1999td}
A.~Pilaftsis,
    Phys.\ Lett.\ B {\bf 471}, 174 (1999) [arXiv:hep-ph/9909485].

\bibitem{Demir:2003js}
    D.~A.~Demir, O.~Lebedev, K.~A.~Olive, M.~Pospelov and A.~Ritz,
    Nucl.\ Phys.\ B {\bf 680} (2004) 339 [arXiv:hep-ph/0311314].

\bibitem{Dedes:1999sj}
A.~Dedes and S.~Moretti,
    Phys.\ Rev.\ Lett.\  {\bf 84} (2000) 22, [hep-ph/9908516];
    Nucl.\ Phys.\ B {\bf 576} (2000) 29, [hep-ph/990941].

\bibitem{Choi:2001iu}
S.~Y.~Choi, K.~Hagiwara and J.~S.~Lee,
    Phys.\ Lett.\ B {\bf 529} (2002) 212 [hep-ph/0110138].


\bibitem{Choi:1998ei}
S.~Y.~Choi, A.~Djouadi, H.~S.~Song and P.~M.~Zerwas,
    Eur.\ Phys.\ J.\ C {\bf 8}, 669 (1999) [hep-ph/9812236].

\bibitem{Choi:1999mv}
S.~Y.~Choi, M.~Guchait, H.~S.~Song and W.~Y.~Song,
     Phys.\ Lett.\ B {\bf 483}, 168 (2000) [hep-ph/9904276].


\bibitem{Kneur:1999nx}
J.~L.~Kneur and G.~Moultaka,
    Phys.\ Rev.\ D {\bf 61}, 095003 (2000) [hep-ph/9907360].


\bibitem{Choi:2000ta}
S.~Y.~Choi, A.~Djouadi, M.~Guchait, J.~Kalinowski, H.~S.~Song and P.~M.~Zerwas,
    Eur.\ Phys.\ J.\ C {\bf 14}, 535 (2000) [hep-ph/0002033].


\bibitem{Choi:2000kt}
S.~Y.~Choi, M.~Guchait, H.~S.~Song and W.~Y.~Song,
     hep-ph/0007276.

\bibitem{Choi:2001ww}
S.~Y.~Choi, J.~Kalinowski, G.~Moortgat-Pick and P.~M.~Zerwas,
    Eur.\ Phys.\ J.\ C {\bf 22}, 563 (2001),
    Addendum-ibid.\ C {\bf 23}, 769 (2002)
    [hep-ph/0108117].

\bibitem{Barger:2001nu}
V.~D.~Barger, T.~Falk, T.~Han, J.~Jiang, T.~Li and T.~Plehn,
    Phys.\ Rev.\ D {\bf 64}, 056007 (2001) [hep-ph/0101106].

\bibitem{Bartl:2003tr}
A.~Bartl, H.~Fraas, O.~Kittel and W.~Majerotto,
    Phys.\ Rev.\ D {\bf 69} (2004) 035007 [hep-ph/0308141].

\bibitem{Bartl:2003gr}
A.~Bartl, T.~Kernreiter and O.~Kittel,
    Phys.\ Lett.\ B {\bf 578}, 341 (2004) [hep-ph/0309340].

\bibitem{Choi:2003pq}
S.~Y.~Choi, M.~Drees, B.~Gaissmaier and J.~Song,
    Phys.\ Rev.\ D {\bf 69}, 035008 (2004) [hep-ph/0310284].

\bibitem{Bartl:2004ut}
A.~Bartl, H.~Fraas, O.~Kittel and W.~Majerotto,
     hep-ph/0402016.

\bibitem{Choi:2004rf}
S.~Y.~Choi, M.~Drees and B.~Gaissmaier,
    hep-ph/0403054.

\bibitem{Bartl:2002hi}
A.~Bartl, T.~Kernreiter and W.~Porod,
    Phys.\ Lett.\ B {\bf 538}, 59 (2002) [hep-ph/0202198].

\bibitem{Bartl:2002uy}
A.~Bartl, K.~Hidaka, T.~Kernreiter and W.~Porod,
    Phys.\ Lett.\ B {\bf 538} (2002) 137 [hep-ph/0204071].


\bibitem{Bartl:2002bh}
A.~Bartl, K.~Hidaka, T.~Kernreiter and W.~Porod,
    Phys.\ Rev.\ D {\bf 66} (2002) 115009 [hep-ph/0207186].

\bibitem{Bartl:2003ck}
A.~Bartl, H.~Fraas, T.~Kernreiter and O.~Kittel,
    Eur.\ Phys.\ J.\ C {\bf 33} (2004) 433 [hep-ph/0306304].

\bibitem{Bartl:2003he}
A.~Bartl, S.~Hesselbach, K.~Hidaka, T.~Kernreiter and W.~Porod,
    Phys.\ Lett.\ B {\bf 573} (2003) 153 [hep-ph/0307317].

\bibitem{Bartl:2003pd}
A.~Bartl, S.~Hesselbach, K.~Hidaka, T.~Kernreiter and W.~Porod,
    hep-ph/0311338.


\bibitem{Choi:2001pg}
S.~Y.~Choi, K.~Hagiwara and J.~S.~Lee,
    Phys.\ Rev.\ D {\bf 64} (2001) 032004 [hep-ph/0103294];

\bibitem{Arhrib:2001pg}
A.~Arhrib, D.~K.~Ghosh and O.~C.~W.~Kong,
        Phys.\ Lett.\ B {\bf 537} (2002) 217 [hep-ph/0112039].

\bibitem{Choi:2002zp}
S.~Y.~Choi, M.~Drees, J.~S.~Lee and J.~Song,
    Eur.\ Phys.\ J.\ C {\bf 25}, 307 (2002) [hep-ph/0204200].


\bibitem{Carena:2002bb}
M.~Carena, J.~R.~Ellis, S.~Mrenna, A.~Pilaftsis and C.~E.~M.~Wagner,
    Nucl.\ Phys.\ B {\bf 659} (2003) 145 [hep-ph/0211467].

\bibitem{Borzumati:2004rd}
F.~Borzumati, J.~S.~Lee and W.~Y.~Song,
    hep-ph/0401024.


\bibitem{Christova:2002ke}
E.~Christova, H.~Eberl, W.~Majerotto and S.~Kraml,
    Nucl.\ Phys.\ B {\bf 639} (2002) 263,
    Erratum-ibid.\ B {\bf 647} (2002) 359 [hep-ph/0205227];
    E.~Christova, H.~Eberl, W.~Majerotto and S.~Kraml,
    JHEP {\bf 0212} (2002) 021 [hep-ph/0211063].


\bibitem{Nojiri:1994it}
M.~M.~Nojiri,
    Phys.\ Rev.\ D {\bf 51} (1995) 6281 [hep-ph/9412374].

\bibitem{Nojiri:1996fp}
M.~M.~Nojiri, K.~Fujii and T.~Tsukamoto,
    Phys.\ Rev.\ D {\bf 54}, 6756 (1996) [hep-ph/9606370].

\bibitem{Boos:2003vf}
E.~Boos, H.~U.~Martyn, G.~Moortgat-Pick,
    M.~Sachwitz, A.~Sherstnev and P.~M.~Zerwas,
    Eur.\ Phys.\ J.\ C {\bf 30} (2003) 395 [hep-ph/0303110].


\bibitem{lepsusy}
LEPSUSYWG, ALEPH, DELPHI, L3 and OPAL experiments,
    note LEPSUSYWG/01-03, {\tt http://lepsusy.web.cern.ch/lepsusy}

\bibitem{Aguilar-Saavedra:2001rg}
J.~A.~Aguilar-Saavedra {\it et al.}
    [ECFA/DESY LC Physics Working Group Collaboration],
    {\it TESLA Technical Design Report Part III: Physics at an $e^+e^-$
         Linear Collider},  hep-ph/0106315.

\bibitem{Bartl:1997yi}
A.~Bartl, H.~Eberl, S.~Kraml, W.~Majerotto, W.~Porod and A.~Sopczak,
    Z.\ Phys.\ C {\bf 76} (1997) 549 [hep-ph/9701336];
A.~Bartl, H.~Eberl, S.~Kraml, W.~Majerotto and W.~Porod,
    Eur.\ Phys.\ J.\ directC {\bf 2} (2000) 6 [hep-ph/0002115].

\bibitem{Belanger:2001am}
See, for instance,
G.~Belanger, F.~Boudjema, A.~Cottrant, R.~M.~Godbole and A.~Semenov,
    Phys.\ Lett.\ B {\bf 519} (2001) 93 [hep-ph/0106275].

\bibitem{Majerotto:2002iu}
    For a review 
    see:
    W.~Majerotto,
    arXiv:hep-ph/0209137,
    and references therein.

\bibitem{Godbole:2004xe}
For a recent summary, see
R.~M.~Godbole, S.~Kraml, M.~Krawczyk, D.~J.~Miller, P.~Niezurawski
    and A.~F.~Zarnecki, 
    hep-ph/0404024.

\bibitem{spa}
SUSY Parameter Analysis (SPA) project,
    {\tt http://www-flc.desy.de/spa/}


\end{thebibliography}
\end{document}